\documentclass{article}

\PassOptionsToPackage{numbers,compress}{natbib}

\usepackage[preprint]{neurips_2025}

\usepackage[utf8]{inputenc}
\usepackage[T1]{fontenc}
\usepackage{microtype}
\usepackage{graphicx}
\usepackage{subcaption}
\usepackage{booktabs}
\usepackage{multirow}
\usepackage{tcolorbox}
\usepackage[table]{xcolor}
\usepackage{hyperref}
\usepackage{url}
\usepackage{nicefrac}

\usepackage{amsmath}
\usepackage{amssymb}
\usepackage{mathtools}
\usepackage{amsthm}
\usepackage{enumitem}
\usepackage{algorithm}
\usepackage{algorithmic}

\usepackage[capitalize,noabbrev]{cleveref}

\theoremstyle{plain}

\theoremstyle{definition}

\theoremstyle{remark}

\newenvironment{squishitemize}
{\begin{list}{\textbullet}{%
    \setlength{\itemsep}{0pt}%
    \setlength{\parsep}{0pt}%
    \setlength{\topsep}{0pt}%
    \setlength{\parskip}{0pt}%
    \setlength{\labelwidth}{.5in}%
    \setlength{\labelsep}{0.05in}%
    \setlength{\leftmargin}{.15in}%
    }}
  {\end{list}}

\usepackage[textsize=tiny]{todonotes}

\newif\ifdraft
\drafttrue

\ifdraft
  \newcommand{\varun}[1]{\textcolor{red}{VC: #1}}
  \newcommand{\zhiyuan}[1]{\textcolor{purple}{ZY: #1}}
  \newcommand{\varunfix}[1]{\textcolor{olive}{VC: #1}}
  \newcommand{\taoran}[1]{\textcolor{blue}{TL: #1}}
\else
  \newcommand{\varun}[1]{}
  \newcommand{\zhiyuan}[1]{}
  \newcommand{\varunfix}[1]{}
  \newcommand{\taoran}[1]{}
\fi

\title{Layer-Targeted Multilingual Knowledge Erasure in Large Language Models}

\author{%
  Taoran Li \\
  Department of Computer Science and Engineering \\
  Texas A\&M University \\
  College Station, TX, US \\
  \And
  Varun Chandrasekaran \\
  Department of Electrical \& Computer Engineering \\
  University of Illinois at Urbana-Champaign \\
  Urbana, IL, US \\
  \texttt{varunc@illinois.edu} \\
  \And
  Zhiyuan Yu \\
  Department of Computer Science and Engineering \\
  Texas A\&M University \\
  College Station, TX, US \\
  \texttt{zhiyuanyu@tamu.edu} \\
}

\begin{document}

\maketitle

\begin{abstract}

Recent work has demonstrated that machine unlearning in Large Language Models (LLMs) fails to generalize across languages: knowledge erased in one language frequently remains accessible through others. However, the underlying cause of this failure and a principled solution remain open. In this work, we identify intervention depth as the key factor determining multilingual generalization. Through systematic layer-wise experiments, we characterize two distinct failure modes: shallow-layer interventions achieve erasure but collapse multilingual capabilities in held-out languages, while deep-layer interventions preserve utility but fail to erase target knowledge even in source languages. These findings reveal that the choice of intervention layer is not a free parameter; it fundamentally determines whether multilingual unlearning succeeds. We propose MUTE (Multilingual Unlearning via Targeted Erasure), a framework that uses Centered Kernel Alignment (CKA) and Linguistic Regions Development Score (LRDS) to identify intermediate, language-agnostic layers where cross-lingual representations converge. By restricting unlearning updates to these layers, MUTE achieves robust multilingual knowledge erasure while optimizing on only a small set of source languages. Extensive experiments across three LLM architectures and three unlearning algorithms validate our approach, with mechanistic analysis via Logit Lens probing confirming genuine knowledge removal rather than output-level suppression.

\end{abstract}
\section{Introduction}

\label{sec:intro}

\begin{figure}[t]
    \centering
    \includegraphics[width=0.95\columnwidth]{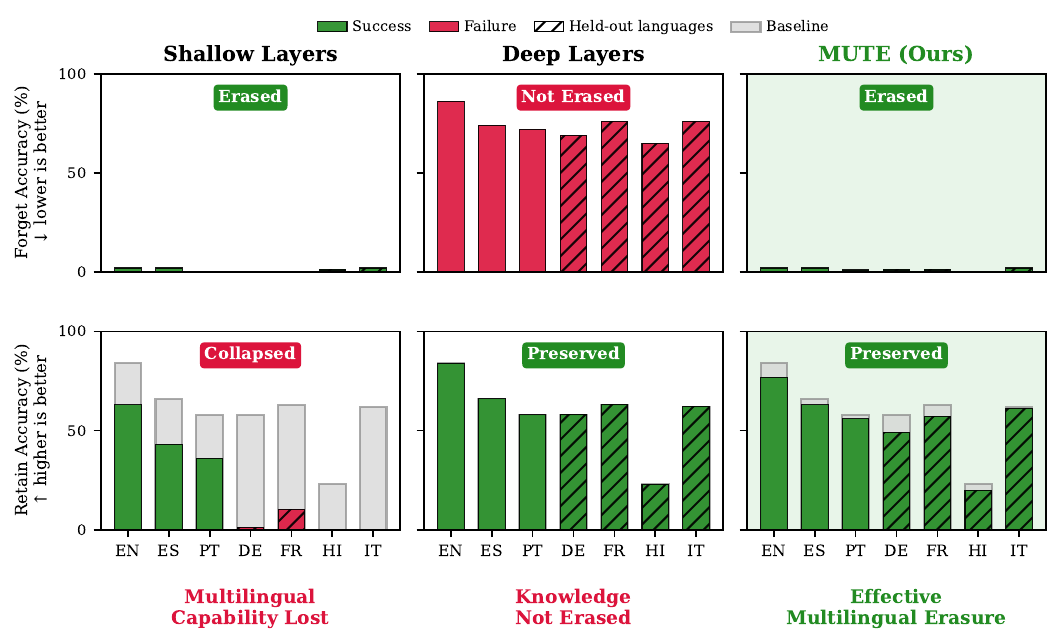}
    \caption{
        \textbf{Multilingual unlearning requires targeting the right depth.}
        We evaluate unlearning interventions at different layer depths, 
        training on 3 source languages (EN, ES, PT) and evaluating on 7 languages including 4 held-out languages (DE, FR, HI, IT).
        \textbf{Left:} Shallow layer interventions effectively erase knowledge but catastrophically collapse multilingual capabilities. Held-out languages lose nearly all utility.
        \textbf{Middle:} Deep layer interventions preserve utility but fail to erase knowledge. The target information remains accessible.
        \textbf{Right:} Our method MUTE targets a language-agnostic region identified via multilingual representation analysis, achieving effective erasure while preserving multilingual utility. \vspace{-4mm}
    }
    \label{fig:teaser}
\end{figure}

Large Language Models (LLMs) increasingly serve as global information infrastructure, supporting hundreds of languages and billions of users worldwide~\citep{nllbteam2022languageleftbehindscaling}. This global reach introduces a critical safety challenge. When hazardous knowledge must be removed from a model, such as instructions for synthesizing dangerous chemicals or exploiting security vulnerabilities, the unlearning must be effective across \emph{all} supported languages, not merely the dominant ones used during safety interventions.

Current machine unlearning approaches fundamentally operate in the monolingual setting. Methods based on gradient ascent~\citep{jang-etal-2023-knowledge}, preference optimization~\citep{zhang2024negativepreferenceoptimizationcatastrophic,fan2025simplicityprevailsrethinkingnegative}, or representation editing~\citep{li2024wmdpbenchmarkmeasuringreducing} typically optimize on English data and implicitly assume the unlearning effect will generalize. However, this assumption proves dangerously flawed in multilingual contexts. Recent work on multilingual jailbreaks demonstrates that safety mechanisms frequently fail to generalize across languages, allowing attackers to bypass English-trained safety filters simply by translating queries into low-resource languages~\citep{wei2023jailbrokendoesllmsafety,deng2024multilingualjailbreakchallengeslarge,yong2024lowresourcelanguagesjailbreakgpt4}. We hypothesize that a parallel vulnerability exists for unlearning: knowledge erased in English may remain fully accessible in other natural languages. 

A naive approach is to curate unlearning datasets for every supported language, but it is computationally prohibitive and impractical. Modern multilingual LLMs support numerous languages, many with limited data availability. A scalable solution must achieve \emph{multilingual generalization}: unlearning on a small subset of source languages should propagate the erasure effect to all held-out languages. 

We argue that achieving such multilingual generalization requires rethinking \emph{where} in the model architecture unlearning interventions should occur. Most existing methods apply updates globally across all layers or target the final output layers, implicitly treating all network depths as equivalent. However, multilingual LLMs exhibit a distinctive internal structure: shallow layers encode language-specific surface features, intermediate layers develop language-agnostic semantic representations, and deep layers re-specialize for language-specific generation~\citep{chi2020finding, wu2019emerging, bhattacharya2023unveilingmultilingualitytransformermodels, wendler2024llamasworkenglishlatent, zhao2024largelanguagemodelshandle, liu2025middlelayerrepresentationalignmentcrosslingual, skean2025layerlayeruncoveringhidden}. This structure suggests that the choice of the intervention layer should fundamentally determine whether multi-unlearning generalizes.

Through systematic experiments varying the intervention layer, we identify two critical failure modes that validate this hypothesis: (1) Interventions at deep layers fail to erase knowledge even on source languages, as representations have diverged into language-specific subspaces; and (2) Interventions at shallow layers achieve erasure but damage multilingual capabilities on held-out languages.

These failure modes reveal that \emph{where} we intervene matters as much as \emph{what} we optimize. We propose the MUTE (\textbf{M}ultilingual \textbf{U}nlearning via \textbf{T}argeted \textbf{E}rasure) framework, which addresses multilingual unlearning through two complementary contributions. First, we develop a principled method for \emph{localizing} the optimal intervention point using Centered Kernel Alignment (CKA)~\citep{kornblith2019similarityneuralnetworkrepresentations} to measure multilingual representational similarity and Linguistic Regions Development Score (LRDS)~\citep{zeng2025converginglinguafrancaevolution} to quantify language specificity. This CKA-LRDS approach identifies an intermediate layer where multilingual representations are maximally aligned and minimally language-specific. Second, we develop \emph{layer-targeted adaptations} of three representative unlearning algorithms (RMU, SLUG, SimNPO) that concentrate parameter updates at the identified layer rather than across the entire network. 

We validate our framework across three model architectures (Llama-3.1, Qwen-2.5, BLOOM). Our experiments demonstrate that layer-targeted interventions achieve near-perfect erasure across multiple languages while preserving general capabilities, {\em all by optimizing on only 3 source languages}. Mechanistic verification via Logit Lens probing~\citep{geva2021transformerfeedforwardlayerskeyvalue,lucki2025adversarialperspectivemachineunlearning} confirms that interventions at this layer remove knowledge from intermediate representations rather than merely suppressing output tokens.

\noindent\textbf{Contributions.} Our main contributions are as follows:
\begin{squishitemize}
    \item We identify and characterize two fundamental failure modes that explain why naive multilingual unlearning fails, demonstrating that intervention layer critically determines multilingual generalization (\S\ref{subsec:motivation_pilot}).
    \item We propose MUTE, a framework that uses a CKA-LRDS analysis to identify the language-agnostic region and select the optimal layer for multilingual unlearning (\S\ref{sec:layer_identification}).
    \item We develop layer-targeted adaptations of three unlearning paradigms (RMU, SLUG, SimNPO) that restrict updates to the target layer, enabling effective erasure across up to 12 languages via only 3 source languages (\S\ref{sec:unlearning_methods}).
    \item We provide mechanistic evidence via Logit Lens probing that interventions at the target layer achieve genuine erasure rather than surface-level output masking (\S\ref{subsec:llama_case_study}).
\end{squishitemize}
\section{Related Work}
\label{sec:related_work}


\noindent\textbf{Machine unlearning for LLMs.}
Machine unlearning aims to erase the influence of a specific subset of the training data ($D_{\text{forget}}$) while preserving general utility on a different subset ($D_{\text{retain}}$)~\citep{7163042, bourtoule2020machineunlearning}. While exact unlearning approaches~\citep{bourtoule2020machineunlearning} are feasible for small models, it is computationally prohibitive for LLMs~\citep{yao2024largelanguagemodelunlearning}. Approximate unlearning methods fall into three paradigms:
(1) \textit{Gradient-based optimization} methods including Gradient Ascent (GA)~\citep{jang-etal-2023-knowledge}, Negative Preference Optimization (NPO)~\citep{zhang2024negativepreferenceoptimizationcatastrophic}, SimNPO~\citep{fan2025simplicityprevailsrethinkingnegative}, and task-specific objectives~\citep{chen2023unlearn, li2024effective};
(2) \textit{Localization and pruning} approaches targeting specific components rather than all weights, including ROME~\citep{meng2023locatingeditingfactualassociations}, MEMIT~\citep{meng2023masseditingmemorytransformer}, SLUG~\citep{cai2025targetedunlearningsinglelayer}, and neuron masking~\citep{pochinkov2024dissecting, liu2024learntoforget}; and
(3) \textit{Representation engineering} methods such as RMU~\citep{li2024wmdpbenchmarkmeasuringreducing} operate on activation space by steering hidden states toward random vectors.


\noindent\textbf{Multilingual safety and representation learning.}
Ensuring safety across languages is critical for LLMs~\citep{wang2024languagesmattermultilingualsafety}. Studies on multilingual jailbreaks show that safety mechanisms trained on high-resource languages often fail when queries are translated into low-resource languages~\citep{wei2023jailbrokendoesllmsafety, deng2024multilingualjailbreakchallengeslarge, yong2024lowresourcelanguagesjailbreakgpt4}.
Understanding multilingual representations is essential for addressing this vulnerability. Research has progressed from embedding alignment~\citep{schuster2019cross, cao2019multilingual} and unsupervised methods~\citep{lian2020unsupervised} to transformer-based multilingual pretraining~\citep{lample2019cross} and analysis of emergent multilingual structures~\citep{wu2019emerging, chi2020finding, artetxe2020cross, bhattacharya2023unveilingmultilingualitytransformermodels, dumas2024separating}. Metrics for quantifying representational alignment include CKA~\citep{kornblith2019similarityneuralnetworkrepresentations}, CCA~\citep{morcos2018insightsrepresentationalsimilarityneural}, and LRDS~\citep{zeng2025converginglinguafrancaevolution}.
Recent work has begun exploring multilingual unlearning~\citep{lu2025learnunlearnaddressingmisinformation, choi2024crosslingualunlearningselectiveknowledge}, and \citet{anonymous2025multilingual} discovered that unlearning in one language degrades retention in others.

\section{Formulation and Core Challenges of Multilingual Unlearning}
\label{sec:motivation}
In this section, we formalize the multilingual unlearning problem. The core challenge is achieving erasure across all supported languages while training on only a few. We demonstrate through experiments that this is fundamentally tied to intervention depth: the layer at which we apply unlearning determines erasure effectiveness across languages.

\subsection{Problem Formulation}


Modern multilingual LLMs are trained on corpora spanning dozens to hundreds of languages, including many low-resource languages with limited data availability~\citep{nllbteam2022languageleftbehindscaling}. While this broad coverage enables global deployment, it also creates a safety challenge: hazardous knowledge (e.g., instructions for synthesizing dangerous chemicals or generating malicious code) must be removed consistently across \emph{all} supported languages. 

Formally, we consider a multilingual LLM $\mathcal{M}$ parameterized by $\theta$, composed of $N$ transformer layers: $\mathcal{M}_\theta = f_N \circ f_{N-1} \circ \dots \circ f_1$. 
Let $\mathbf{h}_{l}(x) \in \mathbb{R}^d$ denote the hidden state at layer $l$ for input $x$. Let $\mathcal{L}_{\text{all}}$ denote the set of all supported languages. We partition this into source languages $\mathcal{L}_{\text{src}} \subset \mathcal{L}_{\text{all}}$ (used for unlearning) and a set of hold-out languages $\mathcal{L}_{\text{hold-out}} = \mathcal{L}_{\text{all}} \setminus \mathcal{L}_{\text{src}}$ (used only for evaluation).

The unlearning task involves two datasets: a \emph{forget set} $D_{\text{forget}}$ containing knowledge to be erased (e.g., chemistry questions that could enable synthesis of dangerous substances), and a \emph{retain set} $D_{\text{retain}}$ containing general knowledge that should be preserved (e.g., history and law). Both datasets exist in all languages $\mathcal{L}_{\text{all}}$, but the unlearning algorithm only accesses data from $\mathcal{L}_{\text{src}}$.

Prior work on multilingual jailbreaks has shown that safety mechanisms trained primarily on English frequently fail when queries are translated into low-resource languages~\citep{wei2023jailbrokendoesllmsafety,deng2024multilingualjailbreakchallengeslarge,yong2024lowresourcelanguagesjailbreakgpt4}, suggesting that unlearning may exhibit similar multilingual brittleness. Recent work confirms this concern: \citet{anonymous2025multilingual} demonstrated that standard unlearning methods, which update parameters globally across all layers, fail to generalize erasure across languages while simultaneously degrading retention (a phenomenon they term \emph{Multilingual Amnesia}). Their findings establish the problem but do not propose a solution. We take their empirical observation as a starting point and hypothesize that this failure stems from applying updates uniformly across all layers, motivating our investigation into \emph{where} in the model to intervene. Another naive solution would optimize unlearning on all supported languages. However, this approach is computationally prohibitive (requiring $\mathcal{O}(|\mathcal{L}_{\text{all}}|)$ times more optimization steps).
In our experiments, we use translated dataset MMMLU~\cite{openai2024mmmlu} to enable rigorous evaluation, but our goal is to develop methods that generalize to held-out languages \emph{without} requiring training data in those languages.

\noindent\textbf{Key Objective.} The objective of our work is to find parameters $\theta'$ such that: (1) the model fails to answer queries from $D_{\text{forget}}$ in \emph{all} languages $\mathcal{L}_{\text{all}}$, and (2) the model maintains performance on $D_{\text{retain}}$ across $\mathcal{L}_{\text{all}}$, while given access to samples only from $\mathcal{L}_{\text{src}}$.

\subsection{Depth of Unlearning Interventions}
\label{subsec:motivation_pilot}

Most existing unlearning methods apply parameter updates across all layers or default to modifying final layers. However, multilingual LLMs exhibit a well-documented layered structure: shallow layers process surface-level linguistic features, intermediate layers form abstract semantic representations, and deep layers specialize in language-specific generation~\citep{chi2020finding, wu2019emerging, bhattacharya2023unveilingmultilingualitytransformermodels, wendler2024llamasworkenglishlatent, zhao2024largelanguagemodelshandle}. However, existing unlearning methods either apply updates across the entire model or select layers without a principled framework to identify the optimal depth for multilingual generalization \cite{li2024wmdpbenchmarkmeasuringreducing}. This raises a natural question: \textit{does the depth at which we apply updates during unlearning affect multilingual generalization?}

To answer this question, we conduct a pilot study by applying the RMU algorithm~\citep{li2024wmdpbenchmarkmeasuringreducing} to Llama-3.1-8B at two distinct depths: a shallow layer (Layer 2) and a deep layer (Layer 30, near the output). We use the \textsc{High School Chemistry} subset of MMMLU~\cite{openai2024mmmlu} as the forget set and the \textsc{History/Law} subset as the retain set. We optimize on three source languages (English, Spanish, Portuguese) and evaluate generalization to four held-out languages (German, French, Hindi, Italian). These seven languages span diverse language families and are all well-supported by Llama-3.1.

\noindent\textit{Failure Mode 1.}
Table~\ref{tab:pilot_study} (right) shows that intervening at Layer 30 does not induce effective unlearning even on the source language: accuracy on the English forget set remains at 86.0\%, similar to the finetuned baseline. This occurs because the target knowledge has already been retrieved and processed by earlier layers; deep layers primarily handle language-specific token generation rather than semantic representation. Therefore, modifying deep layers cannot erase the underlying knowledge, which remains encoded in earlier layers. As a result, the target knowledge remains accessible across other languages, including held-out ones.

\noindent\textit{Failure Mode 2.}
Table~\ref{tab:pilot_study} (left) reveals the other failure at Layer 2. Shallow intervention effectively erases the target knowledge: forget set accuracy drops to near-zero across all languages. However, the model's general multilingual capabilities collapse entirely. The retain set accuracy for held-out languages decreases significantly (German: 58.0\%$\rightarrow$1.0\%, Italian: 62.0\%$\rightarrow$0.0\%), while source languages suffer moderate degradation (English: 84.0\%$\rightarrow$63.0\%). This occurs because shallow layers encode fundamental multilingual representations; disrupting them destroys the model's ability to process non-source languages.


\noindent\textbf{Implications.}
These experiments show that intervention depth is not a free parameter: it fundamentally determines the effectiveness of multilingual unlearning. Applying unlearning on deep layers fails to erase knowledge because semantic representations are largely encoded in earlier layers. In contrast, intervening shallow layers achieves erasure but also ruins multilingual representations, thus degrading utility on held-out languages. This motivates our search for \textit{language-agnostic layers}, which are intermediate layers where the same semantic concept, expressed in different languages, maps to similar hidden representations and is minimally language-specific. We hypothesize that such layers form a contiguous space which we term the \textit{language-agnostic region}, and that selecting an appropriate target layer within this region for intervention can modify the shared semantic representation of a concept, thereby achieving multilingual erasure with minimal utility degradation.

\begin{table}[t]
\centering
\caption{\textbf{Motivating experiments including two failure modes of naive layer selection.} We apply RMU to Llama-3.1 at Layer 2 (shallow) and Layer 30 (deep). \textbf{Left:} Shallow intervention leads to successful erasure but catastrophic retain collapse in held-out languages. \textbf{Right:} Deep intervention causes complete failure to unlearn even source languages. Stars (*) denote source languages. FT = Finetuned, Unl = Unlearned.}
\label{tab:pilot_study}
\vspace{0.5em}
\resizebox{\linewidth}{!}{
\tiny 
\begin{tabular}{l|ccc|ccc||ccc|ccc}
\toprule
& \multicolumn{6}{c||}{\textbf{Layer 2 (Shallow)}} & \multicolumn{6}{c}{\textbf{Layer 30 (Deep)}} \\
\cmidrule(lr){2-7} \cmidrule(lr){8-13}
& \multicolumn{3}{c|}{Forget (\%) $\downarrow$} & \multicolumn{3}{c||}{Retain (\%) $\uparrow$} & \multicolumn{3}{c|}{Forget (\%) $\downarrow$} & \multicolumn{3}{c}{Retain (\%) $\uparrow$} \\
\textbf{Lang} & FT & Unl & $\Delta$ & FT & Unl & $\Delta$ & FT & Unl & $\Delta$ & FT & Unl & $\Delta$ \\
\midrule
en* & 86 & 2 & -84 & 84 & 63 & -21 & 86 & 86 & +0 & 84 & 84 & +0 \\
es* & 74 & 2 & -72 & 66 & 43 & -23 & 74 & 74 & +0 & 66 & 66 & +0 \\
pt* & 74 & 0 & -74 & 58 & 36 & -22 & 74 & 72 & -2 & 58 & 58 & +0 \\
\midrule
de  & 68 & 0 & -68 & 58 &  1 & -57 & 68 & 69 & +1 & 58 & 58 & +0 \\
fr  & 78 & 0 & -78 & 63 & 10 & -53 & 78 & 76 & -2 & 63 & 63 & +0 \\
hi  & 64 & 1 & -63 & 23 &  0 & -23 & 64 & 65 & +1 & 23 & 23 & +0 \\
it  & 80 & 2 & -78 & 62 &  0 & -62 & 80 & 76 & -4 & 62 & 62 & +0 \\
\bottomrule
\end{tabular}
}
\end{table}
\section{Methodology}
\label{sec:methodology}
\textbf{Overview.} The failure modes identified in Section~\ref{subsec:motivation_pilot} suggest that the intervention layer largely affects the generalizability of unlearning across languages. Motivated by this observation, we propose MUTE (Multilingual Unlearning via Targeted Erasure), a two-stage framework that first localizes the optimal intervention layer through principled metric analysis, then applies layer-targeted unlearning exclusively at that layer (Algorithm 1). MUTE first identifies the language-agnostic region via CKA-LRDS analysis (CKA metric for multilingual alignment, LRDS metric for language specificity), then restricts unlearning updates exclusively to the selected target layer within this region.

\begin{table}[t]
\centering
\small
\renewcommand{\arraystretch}{1.15}
\begin{tabular}{@{}p{\columnwidth}@{}}
\toprule
\textbf{Algorithm 1} Multilingual Unlearning via Targeted Erasure \\
\midrule
\textbf{Require:} Multilingual LLM $M_\theta$ with $L$ layers, forget set $D_{\text{forget}}$, retain set $D_{\text{retain}}$, source languages $\mathcal{L}_{\text{src}}$, scaling factor $\alpha$ \\
\textbf{Ensure:} Unlearned model $M_{\theta'}$ \\
\midrule
1: \textbf{// Stage 1: Target Layer Localization} \\
2: \textbf{for} each layer $l \in \{1, \ldots, N\}$ \textbf{do} \\
3: \quad Compute multilingual alignment $\text{Align}_l$ via CKA \\
4: \quad Compute language specificity $\text{LRDS}_l$ \\
5: \textbf{end for} \\
6: $\tau_{\text{align}} \leftarrow \mathbb{E}[\text{Align}_l]$; \quad $\tau_{\text{spec}} \leftarrow \alpha \cdot \min_l \text{LRDS}_l$ \\
7: $\Lambda \leftarrow \{ l \mid \text{Align}_l \ge \tau_{\text{align}} \land \text{LRDS}_l \le \tau_{\text{spec}} \}$ \\
8: Select $l^* \in \Lambda$ based on unlearning algorithm \\
9: \textbf{// Stage 2: Layer-Targeted Unlearning} \\
10: Freeze all parameters $\theta$ except $\theta_{l^*}$ \\
11: Apply unlearning using $D_{\text{forget}}^{\mathcal{L}_{\text{src}}}$ and $D_{\text{retain}}^{\mathcal{L}_{\text{src}}}$ \\
12: Update only $\theta_{l^*} \rightarrow \theta'_{l^*}$ \\
13: \textbf{return} $M_{\theta'}$ \\
\bottomrule
\end{tabular}

\label{alg:mutealgo}
\end{table}

\subsection{Identifying Language Agnostic Region}
\label{sec:layer_identification}

To effectively unlearn target knowledge across languages with minimal cost, we must identify the optimal intervention point within the model. We seek layers where the same semantic concept, regardless of input language, maps to similar hidden state representations. Such layers are language-agnostic, organizing information by semantics rather than linguistic surface forms. To achieve this, we employ two complementary metrics: CKA measures multilingual representational similarity, while LRDS quantifies the degree to which representations cluster by language identity versus semantic content. Together, they identify layers that are both multilingually aligned and minimally language-specific.

\noindent{\textbf{CKA-LRDS Layer Analysis.}}
We employ two complementary metrics to identify language-agnostic layers. First, we use linear CKA~\cite{kornblith2019similarityneuralnetworkrepresentations} to measure multilingual alignment 
(i.e., the degree to which the same semantic concept, expressed in different languages, maps to similar hidden representations): for each layer $l \in [N]$, we compute the average pairwise CKA score across all supported languages to obtain an aggregate alignment score $\text{Align}_l$, where higher values indicate language-invariant representations. Second, we adapt LRDS~\cite{zeng2025converginglinguafrancaevolution} to measure language specificity 
(i.e., the extent to which representations cluster by language identity rather than semantic content), where LRDS quantifies the divergence between intra-language and inter-language similarity, where values near zero indicate semantics-based rather than language-based clustering. Together, these metrics identify layers that are both multilingually aligned and minimally language-specific. Details are provided in Appendix~\ref{app:cka_details} and~\ref{app:lrds_details}.

\noindent{\textbf{Formalized Selection Strategy.}}
We formalize the selection of the target layer $l^*$ as a constrained optimization problem. Specifically, we seek to identify layers that simultaneously exhibit high multilingual alignment and low language specificity, which requires defining thresholds for both metrics. To enable consistent threshold selection across architectures with different absolute metric ranges, we compute model-specific thresholds using the following unified formulas:
\begin{equation}
\small
    \tau_{\text{align}} = \mathbb{E}[\text{Align}_l], \quad \tau_{\text{spec}} = \alpha \cdot \min_l \text{LRDS}_l,
    \label{eq:threshold}
\end{equation}
where $\mathbb{E}[\text{Align}_l]$ denotes the mean CKA alignment across all layers, and $\alpha$ is a scaling factor that controls the LRDS tolerance. The alignment threshold selects layers with above-average multilingual alignment, while the specificity threshold identifies the range where LRDS remains low. The value of $\alpha$ is architecture-dependent ($\alpha \in [1.5, 4.5]$ in our experiments; see Appendix~\ref{app:threshold_sensitivity} for details).

\begin{figure}[t]
    \centering
    \includegraphics[width=\linewidth]{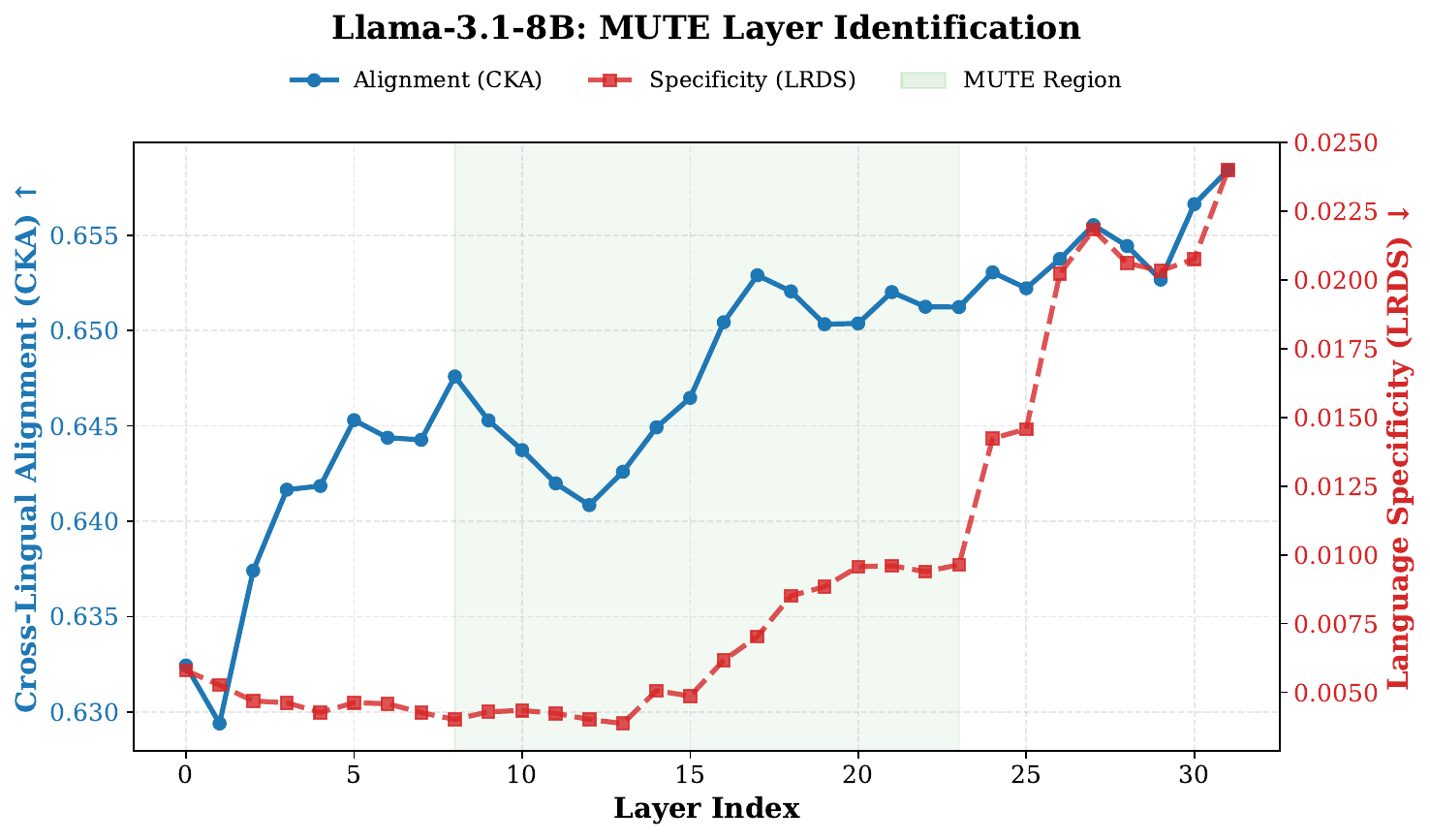}
    \caption{\textbf{Identifying the language-agnostic region.} CKA-LRDS analysis for Llama-3.1. Blue line: multilingual CKA alignment (higher is better). Red line: LRDS (lower is more language-agnostic). Green shaded area: \textit{language-agnostic region} $\Lambda$, satisfying both $\text{CKA} > \tau_{\text{align}}$ and $\text{LRDS} < \tau_{\text{spec}}$, where thresholds are computed via Equation~\ref{eq:threshold}. We select Layer 9 for parameter-based unlearning and Layer 20 for activation-based unlearning.\vspace{-3mm}}
    \label{fig:dual_metric_selection}
    
\end{figure}

As described in~\ref{subsec:motivation_pilot}, we identify the \textit{language-agnostic region} $\Lambda$ as :
\begin{equation}
    \Lambda = \{ l \mid \text{Align}_l \ge \tau_{\text{align}} \land \text{LRDS}_l \le \tau_{\text{spec}} \}.
\end{equation}
For Llama-3.1, applying Equation~\ref{eq:threshold} with $\alpha=2.5$ yields $\tau_{\text{align}}=0.647$ and $\tau_{\text{spec}}=0.0096$, identifying a continuous region $\Lambda = [8, 23]$ (highlighted in green in Figure \ref{fig:dual_metric_selection}). Within these layers, the optimal intervention layer depends on the mechanism of the unlearning algorithm.

\textit{For Parameter-based Methods (e.g., RMU).} These methods modify weights to steer activations away from the knowledge we want to unlearn (target knowledge). 
Intervening early within $\Lambda$ ensures that the disruption propagates through all subsequent layers, preventing the model from reconstructing the target concept downstream. We therefore select the earliest layer where alignment reaches its peak:
\begin{equation}
\small
    l^*_{\text{RMU}} = \min \{ l \in \Lambda \mid \text{Align}_l \approx \max_{k \in \Lambda}(\text{Align}_k) \}.
\end{equation}

\textit{For Activation-based Methods (e.g., SLUG).} These methods require accurate probing of concept representations to compute effective unlearning gradients. Deeper layers within $\Lambda$ encode more abstract, fully-formed semantic representations, improving probe accuracy while remaining language-agnostic. We therefore select the deepest layer within $\Lambda$:
\begin{equation}
\label{eq:slug_layer}
\small
    l^*_{\text{SLUG}} = \max \{ l \in \Lambda \}.
\end{equation}

\subsection{Targeted Unlearning on Language-Agnostic Layers}
\label{sec:unlearning_methods}
Existing unlearning methods are not designed with multilingual generalization in mind. They either operate globally across all layers or select intervention points via task-specific hyperparameter search, without principled justification for why a particular depth should generalize across languages. To validate that our identified target layer enables effective multilingual erasure regardless of the underlying algorithm, we adapt three representative unlearning paradigms: RMU, SLUG, and SimNPO. For each method, we restrict parameter updates to the identified target layer.

\noindent{\bf Adapted Representation Misalignment (RMU).} 
The original RMU~\cite{li2024wmdpbenchmarkmeasuringreducing} selects intervention layers 
via hyperparameter search and updates parameters across three consecutive layers, without accounting for multilingual representation structure. We restrict parameter updates to the single target layer $l^*_{\text{RMU}}$ (e.g., Layer 9 for Llama-3.1) identified via our CKA-LRDS analysis.
We optimize the weights at layer $l^*$ using an $\ell_2$ loss that pushes $\mathbf{h}_{l^*}(x_{\text{forget}})$ towards a random vector $\mathbf{u}$ 
that is fixed throughout training (to ensure a consistent optimization target) and scaled to match the typical magnitude of hidden states. We constrain the representations of retain data to remain close to the frozen base model. By targeting this layer, we disrupt the shared multilingual representation of the target concept before it diverges into language-specific variations, facilitating generalization to held-out languages.

\noindent{\bf Adapted Single Layer Unlearning Gradient (SLUG).} 
SLUG~\cite{cai2025targetedunlearningsinglelayer} performs one-shot unlearning on a single layer selected via Pareto frontier analysis, but its selection criterion optimizes for monolingual forget-retain trade-offs without considering multilingual generalization. We adapt SLUG as a targeted verification procedure for the deep boundary of the proposed language-agnostic region by restricting the gradient computation and parameter update to a specific layer and parameter subset. Unlike the original SLUG which relies solely on the forget gradient, our adaptation incorporates an explicit retain constraint inspired by RMU. Optimization is confined to the deepest layer within the identified language-agnostic region, denoted $l^*_{\text{SLUG}}$ (e.g., Layer 20 in Llama-3.1), which captures high-level semantic representations prior to language-specific specialization (Section~\ref{sec:layer_identification}). 
To balance forgetting and utility preservation, we define the update direction as:
\begin{equation}
\small
\Delta \theta = \nabla_{\theta_{l^*_{\text{SLUG}}}} \mathcal{L}_{\text{forget}} - \alpha \nabla_{\theta_{l^*_{\text{SLUG}}}} \mathcal{L}_{\text{retain}},
\end{equation}
where $\theta_{l^*_{\text{SLUG}}}$ denotes the parameters at layer $l^*_{\text{SLUG}}$. Moving along $\nabla\mathcal{L}_{\text{forget}}$ increases the model's loss on target content, while $-\alpha\nabla\mathcal{L}_{\text{retain}}$ helps preserve utility on general-purpose data, with $\alpha$ controlling this trade-off. 
Parameter updates are applied exclusively to the self-attention matrices $(\mathbf{W}_q, \mathbf{W}_k, \mathbf{W}_v, \mathbf{W}_o)$ at layer $l^*_{\text{SLUG}}$, while all MLP parameters are fixed, constraining the intervention to the model's information-routing mechanisms rather than its stored representations. The resulting update is performed in a single step $\theta_{l^*_{\text{SLUG}}} \leftarrow \theta_{l^*_{\text{SLUG}}} + \lambda \Delta \theta$, where $\lambda$ denotes the step size. 
This procedure tests whether semantic representations at the language-agnostic region boundary can be selectively disrupted to induce global forgetting across languages.

\noindent{\bf Adapted Simple Negative Preference Optimization (SimNPO).}
SimNPO~\cite{fan2025simplicityprevailsrethinkingnegative} frames unlearning as preference optimization with a length-normalized objective that eliminates the reference model required by NPO~\cite{zhang2024negativepreferenceoptimizationcatastrophic}, but updates the entire parameter space.
We hypothesize that modifying the identified language-agnostic layer alone should be sufficient to satisfy the SimNPO unlearning objective. Specifically, we freeze all model parameters $\theta$ except for those in the target layer $l^*_{\text{RMU}}$ (the same layer used for RMU, as both methods perform parameter-based optimization). 
The unlearning gradients are computed at the final output layer but are backpropagated solely to update $\theta_{l^*_{\text{RMU}}}$, testing whether the unlearning signal can be effectively compressed into a single internal layer. We utilize the reference-free SimNPO loss:
\begin{equation}
\small
    \mathcal{L}_{\text{SimNPO}} = - \frac{2}{\beta} \log \sigma \left( -\frac{\beta}{|y|} \log \pi_{\theta_{l^*_{\text{RMU}}}}(y|x) - \gamma \right),
\end{equation}
where $x$ is the input prompt, $y$ is the target response to be unlearned, $\beta$ is a temperature parameter, $\gamma$ is a margin hyperparameter, and $\frac{1}{|y|}$ normalizes the log-probability by sequence length for robustness to varying response lengths across languages. 
\section{Experiments}
\label{sec:results}
\begin{table}[t]
\centering
\caption{\textbf{Cross-model results.} Average Forget and Retain accuracy (\%) across languages for each model. The target layer achieves the best erasure-utility trade-off across all three architectures.}
\label{tab:cross_model_summary}
\resizebox{\linewidth}{!}{
\tiny 
\begin{tabular}{l|cc|cc|cc}
\toprule
\multirow{2}{*}{\textbf{Model}} & \multicolumn{2}{c|}{\textbf{Shallow}} & \multicolumn{2}{c|}{\textbf{Target}} & \multicolumn{2}{c}{\textbf{Deep}} \\
 & Forget $\downarrow$ & Retain $\uparrow$ & Forget $\downarrow$ & Retain $\uparrow$ & Forget $\downarrow$ & Retain $\uparrow$ \\
\midrule
Llama-3.1-8B (L9) & 1.0\% & 21.9\% & \textbf{1.3\%} & \textbf{54.7\%} & 74.0\% & 59.1\% \\
Qwen-2.5-7B (L19) & 0.3\% & 4.9\% & \textbf{15.8\%} & \textbf{13.3\%} & 33.7\% & 14.0\% \\
BLOOM-7b1 (L5) & 0.4\% & 6.0\% & \textbf{0.0\%} & \textbf{6.4\%} & 12.9\% & 7.9\% \\
\bottomrule
\end{tabular}
}
\end{table}
\begin{table*}[t]
\centering
\scriptsize
\caption{\textbf{Behavioral Results (QA Accuracy) on Llama-3.1 using RMU.} We report accuracy (\%) for Forget (Chemistry) and Retain (History/Law) sets across 7 languages. Shallow layers (L2) achieve erasure but degrade Retain performance for held-out languages. Deep layers (L30) fail to erase knowledge. Target layers (L9, L16) within the language-agnostic region achieve effective erasure while preserving utility. * denotes source languages used for optimization.}
\label{tab:behavioral_results}
\begin{tabular}{l|cc|ccccc}
\toprule
\multirow{2}{*}{\textbf{Language}} & \textbf{Base} & \textbf{Finetuned} & \multicolumn{5}{c}{\textbf{Unlearned Accuracy by Intervention Layer}} \\
 & (Pre-trained) & (Target) & L2 (Shallow) & L4 (Shallow) & \textbf{L9 (Target)} & \textbf{L16 (Target)} & L30 (Deep) \\
\midrule
\multicolumn{8}{c}{\cellcolor{gray!10}\textbf{FORGET SET: High School Chemistry} (Lower is Better $\downarrow$)} \\
\midrule
English (en)* & 12.0\% & 86.0\% & 2.0\% & 2.0\% & \textbf{2.0\%} & \textbf{0.0\%} & 86.0\% \\
Spanish (es)* & 7.0\% & 74.0\% & 2.0\% & 2.0\% & \textbf{2.0\%} & \textbf{0.0\%} & 74.0\% \\
Portuguese (pt)* & 3.0\% & 74.0\% & 0.0\% & 0.0\% & \textbf{1.0\%} & \textbf{0.0\%} & 72.0\% \\
German (de) & 5.0\% & 68.0\% & 0.0\% & 0.0\% & \textbf{1.0\%} & \textbf{0.0\%} & 69.0\% \\
French (fr) & 5.0\% & 78.0\% & 0.0\% & 0.0\% & \textbf{1.0\%} & \textbf{0.0\%} & 76.0\% \\
Hindi (hi) & 6.0\% & 64.0\% & 1.0\% & 2.0\% & \textbf{0.0\%} & \textbf{0.0\%} & 65.0\% \\
Italian (it) & 5.0\% & 80.0\% & 2.0\% & 2.0\% & \textbf{2.0\%} & \textbf{0.0\%} & 76.0\% \\
\midrule
\multicolumn{8}{c}{\cellcolor{gray!10}\textbf{RETAIN SET: History \& Law} (Higher is Better $\uparrow$, closer to Finetuned)} \\
\midrule
English (en)* & 5.0\% & 84.0\% & 63.0\% & 75.0\% & \textbf{77.0\%} & \textbf{79.0\%} & 84.0\% \\
Spanish (es)* & 1.0\% & 66.0\% & 43.0\% & 56.0\% & \textbf{63.0\%} & \textbf{64.0\%} & 66.0\% \\
Portuguese (pt)* & 1.0\% & 58.0\% & 36.0\% & 54.0\% & \textbf{56.0\%} & \textbf{59.0\%} & 58.0\% \\
German (de) & 2.0\% & 58.0\% & 1.0\% & 35.0\% & \textbf{49.0\%} & \textbf{56.0\%} & 58.0\% \\
French (fr) & 1.0\% & 63.0\% & 10.0\% & 49.0\% & \textbf{57.0\%} & \textbf{60.0\%} & 63.0\% \\
Hindi (hi) & 0.0\% & 23.0\% & 0.0\% & 6.0\% & \textbf{20.0\%} & \textbf{24.0\%} & 23.0\% \\
Italian (it) & 2.0\% & 62.0\% & 0.0\% & 37.0\% & \textbf{61.0\%} & \textbf{59.0\%} & 62.0\% \\
\bottomrule
\end{tabular}
\end{table*}

We validate MUTE from two complementary perspectives: \textit{behavioral evaluation} (QA accuracy on forget and retain sets) and \textit{mechanistic analysis} (Logit Lens probing of internal representations~\cite{lucki2025adversarialperspectivemachineunlearning, jia2025erasureillusionstresstestinggeneralization} to verify that knowledge is removed rather than suppressed at the output layer). 
Experiments span three model architectures, three unlearning algorithms, and up to 12 languages. Our results show that interventions at the identified language-agnostic layer achieve near-complete erasure across all evaluated languages while preserving model utility, with consistent patterns observed across models and algorithms.

\subsection{Experimental Setup and Baselines}

\textbf{Models.} We evaluate MUTE across three well-established multilingual LLMs with diverse architectures: Llama-3.1-8B~\cite{grattafiori2024llama3herdmodels}, Qwen-2.5-7B~\cite{qwen2025qwen25technicalreport}, and BLOOM-7b1~\cite{workshop2023bloom176bparameteropenaccessmultilingual}. 

\textbf{Languages.} We optimize on three randomly selected source languages (English, Spanish, Portuguese) and evaluate the transferred unlearning effectiveness to held-out languages: 4 for Llama-3.1, 9 for Qwen-2.5, and 5 for BLOOM. 

\textbf{Datasets.} Using the MMMLU~\cite{openai2024mmmlu} and MMLU~\cite{hendrycks2021measuringmassivemultitasklanguage} datasets, we select the subset of \textsc{High School Chemistry} as the forget set (proxy for hazardous knowledge) and that of \textsc{History/Law} as the retain set.
Models are fine-tuned on all domains across supported languages before unlearning.

\textbf{Unlearning Algorithms.} We apply our layer-targeted adaptations (Section~\ref{sec:unlearning_methods}) of three representative methods: RMU~\cite{li2024wmdpbenchmarkmeasuringreducing}, SLUG~\cite{cai2025targetedunlearningsinglelayer}, and SimNPO~\cite{fan2025simplicityprevailsrethinkingnegative}. For all methods, we restrict parameter updates exclusively to the identified target layer. 

\textbf{Evaluation Strategy.} Our experimental design involves two complementary approaches: (1) QA accuracy on forget and retain sets, measuring whether target knowledge is erased while general capabilities are preserved, and (2) Logit Lens probing~\cite{lucki2025adversarialperspectivemachineunlearning, jia2025erasureillusionstresstestinggeneralization}, which inspects internal representations to verify whether knowledge is genuinely removed rather than merely suppressed at the output layer. Since the identified language-agnostic layer varies across architectures, we first conduct an in-depth case study on Llama-3.1 using RMU (Section~\ref{subsec:llama_case_study}), then demonstrate generalizability across models (Section~\ref{subsec:cross_model}) and algorithms (Section~\ref{subsec:cross_algorithm}). The full details are provided in~\ref{app:experimental_details}.

\paragraph{Baselines.} For comparison, we adopt all-layer gradient ascent (GA) as baselines~\citep{anonymous2025multilingual}. We evaluate single-language unlearning (English-only, Hindi-only) and multi-language unlearning with varying learning rates. Our experiments confirm that multilingual transfer is asymmetric~\citep{anonymous2025multilingual}. All-layer GA cannot balance erasure and utility: aggressive settings achieve complete erasure but cause utility collapse, while conservative settings preserve utility but fail to erase knowledge (Appendix~\ref{appendix:single_lang_baselines}). In contrast, our experiments with MUTE (\S\ref{subsec:llama_case_study}) show that it achieves both effective erasure ($\sim$1\% forget accuracy) and utility preservation (55.0\% retain accuracy).

\subsection{Case Study with Llama-3.1}
\label{subsec:llama_case_study}

We conduct an in-depth analysis of Llama-3.1, examining both behavioral outcomes (QA accuracy) and internal mechanisms (Logit Lens probing) across five intervention depths: shallow ($l_2$, $l_4$), target ($l_9$, $l_{16}$), and deep ($l_{30}$).

\subsubsection{Behavioral Results}

Table~\ref{tab:behavioral_results} presents QA accuracy across 7 languages. The results reveal three distinct behavioral patterns corresponding to intervention depth.

\noindent\textbf{Shallow Layers.}
Consistent with our findings in the pilot study (Section~\ref{subsec:motivation_pilot}), interventions at Layer 2 successfully reduce forget set accuracy to near-zero across all languages, but catastrophically degrade retain set performance for held-out languages (e.g., German: $58.0\% \rightarrow 1.0\%$, Italian: $62.0\% \rightarrow 0.0\%$), confirming that shallow layers encode fundamental multilingual representations.

\noindent\textbf{Deep Layers.}
At Layer 30, unlearning fails entirely. The model retains high forget set accuracy similar to the finetuned baseline (English: $86.0\%$, German: $69.0\%$), validating that deep-layer representations have already diverged into language-specific subspaces resistant to unlearning.


\noindent\textbf{Target Layers.}
Layers 9 and 16, identified by our CKA-LRDS analysis as target layers, achieve the desired trade-off. Table~\ref{tab:behavioral_results} shows that Layer 9 reduces forget set accuracy to $\leq 2.0\%$ across all 7 languages while maintaining retain set performance (German: $49.0\%$, Italian: $61.0\%$, compared to finetuned $58.0\%$ and $62.0\%$ respectively). Layer 16 achieves complete erasure ($0\%$ forget accuracy) with even stronger retention (English: $79.0\%$ vs.\ finetuned $84.0\%$). Critically, these results hold for both source languages (used during optimization) and held-out languages (zero-shot transfer), confirming the multilingual generalization of interventions at the target layer.

\paragraph{Source Language Ablations.} We verify that MUTE is not sensitive to the choice of source languages. Using DE/FR/IT as source languages instead of EN/ES/PT, MUTE achieves comparable performance: $\sim$1\% forget accuracy and 54.0\% retain accuracy across all languages, including the now held-out EN/ES/PT. This confirms that MUTE's effectiveness stems from targeting the language-agnostic layer rather than from specific language selection (Appendix~\ref{appendix:source_lang_ablation}).

\subsubsection{Mechanistic Verification via Logit Lens}


Behavioral metrics evaluate end-to-end performance but do not reveal whether knowledge is genuinely removed or merely suppressed at the output layer. We employ Logit Lens probing~\cite{geva2021transformerfeedforwardlayerskeyvalue, lucki2025adversarialperspectivemachineunlearning} to inspect internal representations. Table~\ref{tab:mechanistic_results} in~\ref{app:additional_results} reports the probability assigned to correct answers at intermediate layers, providing a white-box view of the inner mechanism.

\noindent\textbf{Shallow Layer.}
At Layer 2, the internal representations for held-out languages show structural degradation. As shown in Table~\ref{tab:mechanistic_results}, Hindi retain set recall drops from $30.05\%$ (finetuned) to $3.47\%$, and German drops from $30.71\%$ to $12.77\%$. This confirms that shallow interventions do not selectively remove target knowledge; instead, they damage the representational capacity required to encode any concept in non-source languages.

\noindent\textbf{Deep Layers.}
At Layer 30, forget set recall remains similar to the finetuned baseline. Table~\ref{tab:mechanistic_results} shows that English recall stays at $29.39\% \to 29.42\%$ and German at $27.44\% \to 27.41\%$. The knowledge persists in the model despite optimization efforts. This explains the behavioral failure: deep-layer interventions do not modify the stored knowledge, rendering them ineffective for unlearning.

\noindent\textbf{Target Layers.}
At Layers 9 and 16, we observe the target mechanistic signature. As Table~\ref{tab:mechanistic_results} demonstrates, forget set recall drops to near-zero (English: $1.50\%$, Hindi: $1.18\%$), confirming that the target knowledge is removed from internal representations. Additionally, retain set recall remains close to the finetuned baseline (Hindi: $29.64\%$ vs.\ $30.05\%$; Italian: $31.32\%$ vs.\ $31.44\%$). This demonstrates that interventions at the target layer achieve knowledge removal while preserving the multilingual representation space.

\subsection{Results on Other Model Architectures}
\label{subsec:cross_model}

To verify that the language-agnostic layer phenomenon is not specific to Llama-3.1, we evaluate two additional architectures: Qwen-2.5-7B and BLOOM-7b1. Results are summarized in Table~\ref{tab:cross_model_summary}, with full per-language breakdowns in Table~\ref{tab:qwen_full} and Table~\ref{tab:bloom_full} in~\ref{app:additional_results}.

\noindent\textbf{Qwen-2.5-7B.}
Our CKA-LRDS analysis 
identifies Layer 19 as the target layer for Qwen-2.5 (Figure~\ref{fig:dual_metric_qwen}). As in Table~\ref{tab:qwen_full}, at Layer 2 (shallow), Forget accuracy drops to near-zero but Retain performance degrades substantially (German: $13.0\% \to 3.0\%$). At Layer 23 (deep), unlearning fails (English Forget: $45.0\%$ vs.\ finetuned $52.0\%$). At Layer 19 (target), Forget accuracy is reduced to $11.0$--$26.0\%$ across languages while Retain performance is largely preserved. Although the erasure is less effective than Llama-3.1, the relative pattern remains consistent: shallow layers degrade utility, deep layers fail to erase, and target layers within the language-agnostic region achieve the best trade-off.

\noindent\textbf{BLOOM-7b1.}
For BLOOM, our analysis identifies Layer 5 as the target layer (Figure~\ref{fig:dual_metric_bloom}). Table~\ref{tab:bloom_full} shows that Layer 5 achieves nearly complete erasure while preserving Retain performance (English: $13.0\%$ vs.\ finetuned $16.0\%$). In contrast, Layer 29 (deep) shows no unlearning effect (English Forget: $18.0\%$ vs.\ finetuned $17.0\%$). Notably, the target layer in BLOOM (Layer 5/30 $\approx$ 17\% depth) is earlier than in Llama-3.1 (Layer 9/32 $\approx$ 28\% depth). This suggests that the optimal intervention point varies across architectures.

\begin{table}[t]
\centering
\caption{\textbf{Cross-algorithm results on Llama-3.1.} Average Forget and Retain accuracy (\%) across all 7 languages. Target layers vary by algorithm: L9 for RMU/SimNPO, L20 for SLUG. SimNPO achieves the best utility preservation.}
\label{tab:cross_algo_summary}
\resizebox{\linewidth}{!}{
\tiny 
\begin{tabular}{l|cc|cc|cc}
\toprule
\multirow{2}{*}{\textbf{Algorithm}} & \multicolumn{2}{c|}{\textbf{Shallow (L2/L3)}} & \multicolumn{2}{c|}{\textbf{Target}} & \multicolumn{2}{c}{\textbf{Deep (L30)}} \\
 & Forget $\downarrow$ & Retain $\uparrow$ & Forget $\downarrow$ & Retain $\uparrow$ & Forget $\downarrow$ & Retain $\uparrow$ \\
\midrule
RMU (L9) & 1.0\% & 21.9\% & \textbf{1.3\%} & \textbf{54.7\%} & 74.0\% & 59.1\% \\
SLUG (L20) & 0.0\% & 0.0\% & \textbf{35.4\%} & \textbf{31.4\%} & 75.0\% & 59.7\% \\
SimNPO (L9) & 0.7\% & 55.3\% & \textbf{0.0\%} & \textbf{58.6\%} & 68.9\% & 58.1\% \\
\bottomrule
\end{tabular}
}
\end{table}

\subsection{Results on Other Unlearning Algorithms}
\label{subsec:cross_algorithm}

We validate that the identified target layer is effective across unlearning methods by evaluating on SLUG (gradient-based, one-shot) and SimNPO (preference-based) on Llama-3.1. Results are summarized in Table~\ref{tab:cross_algo_summary}, with full per-language results in Table~\ref{tab:slug_full} and Table~\ref{tab:simnpo_full} in~\ref{app:additional_results}.

\noindent\textbf{SLUG.}
As shown in Table~\ref{tab:slug_full}, at Layer 9 (target), SLUG achieves complete erasure ($0\%$ Forget accuracy). However, SLUG at this layer also reduces Retain accuracy to $0\%$, indicating that the single-step gradient update is too aggressive when confined to a single layer. At intermediate depths (L18, L20), partial erasure is achieved (Forget: $24.0\%$--$53.0\%$) with moderate Retain preservation. At Layer 30 (deep), unlearning fails entirely (English Forget: $88.0\%$). These results suggest that while SLUG benefits from targeting within the language-agnostic region for erasure, additional regularization is needed to preserve utility.

\noindent\textbf{SimNPO.}
Table~\ref{tab:simnpo_full} shows that SimNPO achieves the most favorable trade-off at the target layer. At Layer 9, SimNPO reduces Forget accuracy to $0\%$ across all languages while maintaining high Retain performance (English: $86.0\%$ vs.\ finetuned $84.0\%$; Spanish: $75.0\%$ vs.\ $66.0\%$). The length-normalized preference objective provides implicit regularization that prevents utility degradation. At Layer 30 (deep), unlearning fails (English Forget: $81.0\%$), consistent with the results of RMU and SLUG.

Across all three algorithms, the observation and insights remain consistent: interventions at the target layer enable effective multilingual knowledge erasure, while deep-layer interventions fail. The choice of algorithm affects the erasure-utility trade-off, with SimNPO providing the best balance.

\section{Conclusion}
\label{sec:conclusion}

We identify a fundamental limitation of existing unlearning methods: knowledge erased in one language often remains accessible through other languages. Through systematic experiments, we demonstrate that intervention depth critically affects multilingual generalization. Specifically, deep-layer interventions fail to erase knowledge, while shallow-layer interventions ruin multilingual capabilities. Based on these observations, we propose MUTE, a framework that localizes and targets language-agnostic layers for unlearning. 
Experiments across three model architectures, three unlearning algorithms, and up to 12 languages show that MUTE achieves effective multilingual erasure while training on only 3 source languages. We further conduct a mechanistic analysis to validate that MUTE achieves knowledge removal rather than output-level suppression. Additional discussion is provided in~\ref{sec:discussion}. In this study, we introduce the paradigm shift of multilingual unlearning from \textit{what data to optimize on} to \textit{where in the model to intervene}.

\newpage
\bibliographystyle{abbrvnat}
\bibliography{references}

\newpage
\clearpage
\appendix
\onecolumn
\renewcommand{\thesection}{Appendix \Alph{section}}
\renewcommand{\thesubsection}{\Alph{section}.\arabic{subsection}}

\section{Layer Identification: Technical Details}
\label{app:layer_identification}

\paragraph{Dataset for Layer Analysis.} We compute CKA and LRDS using the MMLU and MMMLU High School Chemistry subset across all source languages.

\subsection{CKA-based Multilingual Alignment}
\label{app:cka_details}

We quantify the representational similarity between different languages using Linear Centered Kernel Alignment (CKA)~\cite{kornblith2019similarityneuralnetworkrepresentations}. Unlike simple correlation metrics, CKA is invariant to orthogonal transformations and isotropic scaling, making it robust for comparing high-dimensional neural representations.

For a specific layer $l$, let $\mathbf{H}^{\ell}_l \in \mathbb{R}^{N \times d}$ denote the matrix of hidden states for a set of $N$ aligned concepts in language $\ell$, where each hidden state is the last token representation of the input sequence. The multilingual alignment score between two languages $\ell_j$ and $\ell_k$ at layer $l$ is defined as:
\begin{equation}
\text{CKA}_l(\ell_j, \ell_k) = \frac{\text{HSIC}(K, L)}{\sqrt{\text{HSIC}(K, K) \cdot \text{HSIC}(L, L)}}
\end{equation}
where $K = H_l^{\ell_j}(H_l^{\ell_j})^T$ and $L = H_l^{\ell_k}(H_l^{\ell_k})^T$ are the Gram matrices, and $\text{HSIC}$ denotes the Hilbert-Schmidt Independence Criterion.

A higher CKA score indicates that the model represents the same concepts similarly, regardless of the input language. We define the aggregate alignment score for layer $l$ as the average pairwise CKA across all supported languages $\mathcal{L}$:
\begin{equation}
\text{Align}_l = \frac{2}{|\mathcal{L}|(|\mathcal{L}|-1)} \sum_{j<k} \text{CKA}_l(\ell_j, \ell_k)
\end{equation}

The optimal layer candidate via CKA maximizes this alignment:
\begin{equation}
l_{\text{CKA}}^* = \arg\max_{l \in \{1,...,L\}} \text{Align}_l
\end{equation}

\subsection{LRDS-based Language-Agnosticism}
\label{app:lrds_details}

To measure the degree of language-specificity, we adapt the Linguistic Regions Development Score (LRDS) proposed by~\cite{zeng2025converginglinguafrancaevolution}.

For a given layer $l$, we compute the normalized hidden state representation $a_l^s = \text{normalize}(\bar{h}_{l}(s))$ for an input sequence $s$, where $\bar{h}_{l}(s)$ represents the token-averaged hidden state. The semantic similarity between two samples $s_p$ and $s_q$ is given by their cosine similarity $S_l(s_p, s_q) = a_l^{s_p} \cdot a_l^{s_q}$.

LRDS quantifies the divergence between intra-language similarity and inter-language similarity:
\begin{align}
\text{LRDS}_l &= \mathbb{E} \left[ S_l(s_p, s_q) \mid \substack{\text{lang}(p) = \text{lang}(q) \\ \text{sem}(p) \neq \text{sem}(q)} \right] \nonumber \\
&\quad - \mathbb{E} \left[ S_l(s_p, s_q) \mid \substack{\text{lang}(p) \neq \text{lang}(q) \\ \text{sem}(p) \neq \text{sem}(q)} \right]
\end{align}
where $\text{sem}(s)$ denotes the semantic content (Fact ID). A high positive $\text{LRDS}_l$ indicates that representations are clustered by language (language-specific), while a value near zero or negative implies clustering by semantics (language-agnostic).

The optimal language-agnostic layer via LRDS minimizes this score:
\begin{equation}
l_{\text{LRDS}}^* = \arg\min_{l \in \{1,...,L\}} \text{LRDS}_l
\end{equation}

\subsection{Threshold Selection and Sensitivity}
\label{app:threshold_sensitivity}

As described in Section~\ref{sec:layer_identification}, we identify the language-agnostic region $\Lambda$ by selecting layers that simultaneously exhibit high cross-lingual alignment (high CKA) and low language specificity (low LRDS). We formalize this selection using the following threshold formulas:
\begin{equation}
    \tau_{\text{align}} = \mathbb{E}[\text{Align}_l], \quad \tau_{\text{spec}} = \alpha \cdot \min_l \text{LRDS}_l
    \label{eq:threshold_appendix}
\end{equation}

The alignment threshold $\tau_{\text{align}}$ is set to the mean CKA score, selecting layers with above-average cross-lingual alignment. The specificity threshold $\tau_{\text{spec}}$ uses a scaling factor $\alpha$ applied to the minimum LRDS value, capturing the plateau region before language-specific processing intensifies. Table~\ref{tab:threshold_values} summarizes the computed values for each architecture.

\begin{table}[h]
\centering
\caption{Threshold values and language-agnostic regions across architectures.}
\label{tab:threshold_values}
\begin{tabular}{lcccccc}
\toprule
\textbf{Model} & $\mathbb{E}[\text{CKA}]$ & $\min(\text{LRDS})$ & $\alpha$ & $\tau_{\text{spec}}$ & $\Lambda$ \\
\midrule
Llama-3.1-8B & 0.647 & 0.0039 & 2.5 & 0.0096 & Layers 8--23 \\
BLOOM-7b1 & 0.642 & 0.0017 & 4.4 & 0.0073 & Layers 5--22 \\
Qwen-2.5-7B & 0.512 & 0.0287 & 1.5 & 0.0441 & Layers 19--24 \\
\bottomrule
\end{tabular}
\end{table}

\noindent\textbf{Choice of $\alpha$.} The scaling factor $\alpha$ is architecture-dependent, reflecting differences in how LRDS evolves across depth. For Llama-3.1, LRDS remains stable around its minimum through the middle layers before surging after Layer 23; we use $\alpha=2.5$ to capture this plateau. BLOOM exhibits a similar pattern but with a more gradual LRDS increase, requiring a larger $\alpha=4.4$ to include the extended low-LRDS region through Layer 22. For Qwen-2.5, the LRDS distribution is notably flatter with less pronounced variation across layers. Here, the primary constraint becomes the CKA alignment criterion combined with the characteristic LRDS surge beginning at Layer 24; we use $\alpha=1.5$ accordingly. In practice, we recommend visualizing the CKA-LRDS curves and selecting $\alpha$ to capture the stable low-LRDS plateau before the characteristic surge, as illustrated in Figures~\ref{fig:dual_metric_selection}, \ref{fig:dual_metric_qwen}, and \ref{fig:dual_metric_bloom}.

\noindent\textbf{Sensitivity Analysis.} Table~\ref{tab:alpha_sensitivity} demonstrates the robustness of our selection. Varying $\alpha$ by $\pm 0.5$ from the chosen values produces modest changes in region boundaries, but the selected target layers (Layer 9 for Llama, Layer 5 for BLOOM, Layer 19 for Qwen) consistently remain within $\Lambda$ across all tested values.

\begin{table}[h]
\centering
\caption{Sensitivity of the language-agnostic region $\Lambda$ to the scaling factor $\alpha$.}
\label{tab:alpha_sensitivity}
\begin{tabular}{lccc}
\toprule
\textbf{Model} & $\alpha - 0.5$ & $\alpha$ (used) & $\alpha + 0.5$ \\
\midrule
Llama-3.1-8B & Layers 8--17 & Layers 8--23 & Layers 8--23 \\
BLOOM-7b1 & Layers 5--21 & Layers 5--22 & Layers 5--22 \\
Qwen-2.5-7B & Layers 19--21 & Layers 19--24 & Layers 19--26 \\
\bottomrule
\end{tabular}
\end{table}

\clearpage

\section{Experimental Setup: Full Details}
\label{app:experimental_details}

\subsection{Models}

We select three representative multilingual LLMs to test the generalizability of our findings across different positional embeddings and pre-training distributions. Llama-3.1-8B~\cite{grattafiori2024llama3herdmodels} serves as our primary model for analysis, utilizing Rotary Positional Embeddings (RoPE) and exhibiting strong general-purpose multilingual capabilities. Qwen-2.5-7B~\cite{qwen2025qwen25technicalreport} represents models explicitly optimized for multilingual heavy-tail distributions, allowing us to test whether language-agnostic layers exist in models with denser non-English pre-training data. BLOOM-7b1~\cite{workshop2023bloom176bparameteropenaccessmultilingual} represents the ALiBi (Attention with Linear Biases) architecture, enabling us to verify whether the language-agnostic layer phenomenon is architecture-agnostic. 

\subsection{Model Preparation (Domain Knowledge Injection)}
\label{app:model_prep}

Prior to unlearning, we establish a knowledgeable baseline by fine-tuning the base models. We employ the LLaMA-Factory framework~\cite{zheng2024llamafactoryunifiedefficientfinetuning} to perform supervised fine-tuning (SFT) with LoRA~\cite{hu2021loralowrankadaptationlarge}.

The fine-tuning is conducted on all supported languages for each model to ensure that the knowledge is robustly embedded in the model's parameters across all linguistic modes (encompassing English, Chinese, Swahili, etc.). We construct the fine-tuning corpus by sourcing data from the MMLU~\cite{hendrycks2021measuringmassivemultitasklanguage} and MMMLU dataset~\cite{openai2024mmmlu}, which provides multilingual versions of the specific subjects designated for our Forget and Retain sets. By fine-tuning on both the target hazardous domain and control humanities domains across languages, we ensure the model possesses strong, verifiable knowledge in these specific areas before intervention. For hyperparameters, we use LoRA with rank 8 applied to all linear layers, learning rate $5\text{e-}5$ with cosine scheduler, warmup ratio 0.1, effective batch size 16, and train for 2 epochs with bf16 precision.

\subsection{Datasets and Multilingual Split}
\label{app:datasets}

We base our experiments on the subject taxonomy of the Massive Multitask Language Understanding (MMLU) benchmark~\cite{hendrycks2021measuringmassivemultitasklanguage}. However, as the standard MMLU contains only English samples, we employ the MMMLU dataset~\cite{openai2024mmmlu} for all multilingual training and evaluation steps. MMMLU provides professionally translated versions of MMLU subjects, enabling us to simulate the unlearning task across diverse linguistic contexts.

To ensure reproducibility and safety, we employ Chemistry as a proxy for hazardous dual-use knowledge, while using History and Law to represent general safe knowledge. The specific sample counts per language are as follows: the hazardous domain ($D_1$) consists of High School Chemistry ($N=203$ samples/lang), while the safe domains ($D_{2-5}$) include High School World History ($N=237$), Professional Law ($N=1534$), International Law ($N=121$), and Jurisprudence ($N=108$).

Based on these domains, we construct three distinct data partitions for different stages of the pipeline. The fine-tuning set is used for knowledge injection; to ensure the knowledge is robustly embedded, this set comprises all 5 domains ($D_1 \cup \dots \cup D_5$) multiplied by all supported languages (e.g., $\times 7$ for Llama-3.1). The forget set ($D_{\text{forget}}$) is used for unlearning optimization and consists strictly of the hazardous domain ($D_1$) sourced only from the three source languages (En, Es, Pt), yielding a total of $203 \times 3 = 609$ optimization samples. The retain set ($D_{\text{retain}}$) is used for unlearning constraints (e.g., RMU/SLUG retain loss) and aggregates the four safe domains ($D_2 \cup D_3 \cup D_4 \cup D_5$) sourced only from the three source languages (En, Es, Pt), yielding approximately $2000 \times 3$ samples to preserve general capabilities.

We use the same data (MMLU test split and its multilingual translations in MMMLU) for fine-tuning, layer identification, and evaluation. This does not constitute data leakage because our goal is to measure the reduction in accuracy after unlearning. The fine-tuned model achieves high accuracy by design, and we evaluate whether unlearning successfully degrades this accuracy across languages.

\subsection{Language Split}

To rigorously test zero-shot multilingual transfer, we partition the languages into source languages and held-out languages. The source languages ($\mathcal{L}_{\text{src}}$) consist of English, Spanish, and Portuguese, which are used for gradient computation in both $D_{\text{forget}}$ and $D_{\text{retain}}$. The held-out languages ($\mathcal{L}_{\text{held}}$) are explicitly excluded from optimization, and evaluation on these languages measures zero-shot transfer. Specifically, for Llama-3.1, we use 4 held-out languages (German, French, Hindi, Italian); for Qwen-2.5, we use 9 held-out languages (Arabic, German, French, Hindi, Indonesian, Italian, Japanese, Korean, Chinese); and for BLOOM, we use 5 held-out languages (Arabic, French, Hindi, Indonesian, Chinese).

\subsection{Unlearning Algorithm Hyperparameters}
\label{app:hyperparameters}

\noindent\textbf{Hyperparameter Selection.} To ensure fair comparison across intervention depths, we use identical hyperparameters for all layers within each algorithm. This controlled setup ensures that observed performance differences come from the intervention location rather than hyperparameter choices. We selected hyperparameters based on validation performance at the target layer and applied them uniformly across all depths. To verify that the deep-layer failure is due to the layer location itself rather than suboptimal hyperparameters, we conducted additional experiments varying learning rate, regularization strength, and training steps. The failure mode at deep layers persisted across all tested configurations, confirming that deep layers are fundamentally unsuitable for multilingual unlearning regardless of optimization settings.

We evaluate three distinct unlearning paradigms, and for all methods, we apply the MUTE-targeted adaptation described in Section~\ref{sec:unlearning_methods}. For RMU (parameter-based), we apply it to Llama-3.1 (Layer 9), Qwen-2.5 (Layer 19), and BLOOM (Layer 5), using learning rate $1\text{e-}5$ for 500 steps. For SLUG (activation-based), we apply it to Llama-3.1 (Layer 20) with step size $\lambda=32$ and retain constraint weight $\alpha=1$. For SimNPO (preference-based), we apply it to Llama-3.1 (Layer 9) with learning rate $5\text{e-}5$, $\beta=1.0$, $\gamma=1.0$, and length normalization enabled.

\subsection{Evaluation Protocols}
\label{app:eval_protocols}

We employ a rigorous two-tiered evaluation framework, combining standard behavioral metrics with mechanistic probing to verify that unlearning occurs at the representational level rather than merely suppressing output tokens. 

\subsubsection{Behavioral Metrics (Black-box)}

Following the standard protocol for MMMLU evaluation, we quantify the model's end-to-end performance based on the final output distribution. Unlearning Efficacy (UE) is defined as the reduction in accuracy on the Forget Set ($D_{\text{forget}}$) relative to the fine-tuned baseline. To distinguish between optimization success and generalization, we report this in two distinct scopes: UE-Source measures efficacy on the optimization languages ($\mathcal{L}_{\text{src}}$), indicating how well the algorithm minimizes the specific loss function, while UE-Transfer measures efficacy on the held-out languages ($\mathcal{L}_{\text{held}}$), indicating the zero-shot multilingual transfer of the unlearning effect. Multilingual Integrity (MI) is defined as the average accuracy retention on the Retain Set ($D_{\text{retain}}$) across all evaluated languages; high MI indicates that the model preserves its reasoning capabilities in non-target domains (History/Law) and avoids catastrophic forgetting.

\subsubsection{Mechanistic Validation (White-box)}

To confirm our hypothesis that the intervention successfully targets the language-agnostic layer, we employ Logit Lens probing~\cite{lucki2025adversarialperspectivemachineunlearning}. This technique allows us to inspect the model's internal representation of the target concept at intermediate layers.

For a given input $x$ and target answer $y$, we decode the hidden states $h_l$ at layer $l$ directly into the vocabulary space using the pre-trained embedding matrix $E$:
\begin{equation}
    P_{\text{lens}}(y|x, l) = \text{softmax}(E \cdot \text{LayerNorm}(h_l))_y
\end{equation}
We specifically monitor the probability assigned to the correct answer at two critical checkpoints: the target layer (Layer 9) to verify whether the semantic representation of the knowledge has been erased, and the output layer (Layer 32) to verify the final model behavior. A successful unlearning intervention should demonstrate a significant probability drop at the language-agnostic layer, confirming that the knowledge is removed from intermediate representations rather than merely suppressed at the output.


\clearpage

\section{Additional Experimental Results}
\label{app:additional_results}

The first subsection is the detailed results and discussion for the baseline experiments. Then we provide comprehensive experimental results across different model architectures (Qwen-2.5, BLOOM) and unlearning algorithms (SLUG, SimNPO). The table format follows the main paper: bold columns indicate the identified target layer, and asterisks (*) denote source languages ($\mathcal{L}_{\text{src}}$) used for optimization.

\subsection{Single-Language Unlearning Baselines}
\label{appendix:single_lang_baselines}

Following \citet{anonymous2025multilingual}, we investigate whether all-layer gradient ascent (GA) can achieve effective multilingual unlearning. We adapt their methodology to our experimental setting (Llama-3.1-8B, MMLU and MMMLU dataset) and evaluate four baseline configurations using all-layer GA unlearning: (1) Hindi-only, (2) English-only, and (3-4) source languages (EN/ES/PT) with different learning rates.

\begin{table}[h]
\centering
\caption{Single-language GA unlearning baselines on Llama-3.1-8B. We report forget and retain accuracy for each language. * denotes source languages used in MUTE.}
\label{tab:single_lang_baselines}
\resizebox{\textwidth}{!}{
\begin{tabular}{l|ccccccc|c|ccccccc|c}
\toprule
& \multicolumn{8}{c|}{\textbf{Forget Set Accuracy $\downarrow$}} & \multicolumn{8}{c}{\textbf{Retain Set Accuracy $\uparrow$}} \\
\textbf{Method} & EN* & ES* & PT* & DE & FR & HI & IT & Avg & EN* & ES* & PT* & DE & FR & HI & IT & Avg \\
\midrule
Finetuned & 86.0\% & 74.0\% & 74.0\% & 68.0\% & 78.0\% & 64.0\% & 80.0\% & 75.0\% & 84.0\% & 66.0\% & 58.0\% & 58.0\% & 63.0\% & 23.0\% & 62.0\% & 59.0\% \\
\midrule
GA (Hindi-only) & 86.0\% & 74.0\% & 74.0\% & 68.0\% & 78.0\% & 2.0\% & 80.0\% & 66.0\% & 84.0\% & 66.0\% & 58.0\% & 58.0\% & 63.0\% & 11.0\% & 62.0\% & 57.0\% \\
\midrule
GA (English-only) & 28.0\% & 55.0\% & 43.0\% & 48.0\% & 49.0\% & 48.0\% & 49.0\% & 46.0\% & 89.0\% & 60.0\% & 58.0\% & 52.0\% & 59.0\% & 24.0\% & 57.0\% & 57.0\% \\
\midrule
GA (EN/ES/PT, lr=1e-4) & 0.0\% & 0.0\% & 1.0\% & 2.0\% & 1.0\% & 1.0\% & 1.0\% & 1.0\% & 43.0\% & 26.0\% & 16.0\% & 14.0\% & 11.0\% & 6.0\% & 24.0\% & 20.0\% \\
\midrule
GA (EN/ES/PT, lr=2e-5) & 62.0\% & 49.0\% & 36.0\% & 61.0\% & 65.0\% & 59.0\% & 60.0\% & 56.0\% & 88.0\% & 73.0\% & 67.0\% & 58.0\% & 63.0\% & 23.0\% & 60.0\% & 62.0\% \\
\midrule
\textbf{MUTE (Ours)} & \textbf{2.0\%} & \textbf{2.0\%} & \textbf{1.0\%} & \textbf{1.0\%} & \textbf{1.0\%} & \textbf{0.0\%} & \textbf{2.0\%} & \textbf{1.0\%} & \textbf{77.0\%} & \textbf{63.0\%} & \textbf{56.0\%} & \textbf{49.0\%} & \textbf{57.0\%} & \textbf{20.0\%} & \textbf{61.0\%} & \textbf{55.0\%} \\
\bottomrule
\end{tabular}
}
\end{table}

The results in Table~\ref{tab:single_lang_baselines} reveal several important findings about the limitations of single-language GA unlearning:

\textbf{Multilingual generalization is asymmetric.} \citet{anonymous2025multilingual} demonstrated that unlearning in high-resource languages tends to be more stable with stronger cross-lingual transfer, while low-resource language unlearning exhibits weaker propagation. Our comparison between English-only and Hindi-only GA confirms this asymmetry. English-only GA achieves partial transfer to other languages (16-31\% decrease across non-English languages), whereas Hindi-only GA produces zero transfer and all other languages remain completely unchanged. This validates that high-resource language unlearning generalizes better, while low-resource language unlearning remains isolated.

\textbf{Single-language GA fails to achieve comprehensive erasure.} Despite the partial transfer observed with English-only GA, it remains insufficient for complete multilingual unlearning. English accuracy drops significantly (86.0\%$\rightarrow$28.0\%), but other languages retain substantial knowledge (43.0\%-55.0\% accuracy). Hindi-only GA is even more limited, achieving erasure only in Hindi (64.0\%$\rightarrow$2.0\%) with no effect on other languages.

\textbf{Aggressive GA causes utility collapse.} When we increase the learning rate to achieve complete multilingual erasure (EN/ES/PT, lr=1e-4), the model successfully reduces forget accuracy to $\sim$1\% across all languages. However, this comes at the cost of utility degradation with retain accuracy dropping from 59.0\% to 20.0\% on average.

\textbf{Conservative GA fails to achieve erasure.} With a lower learning rate (lr=2e-5), GA preserves utility (62\% retain accuracy) but fails to achieve effective erasure, with forget accuracy remaining at 36.0-65.0\% across languages.

\textbf{MUTE achieves both erasure and utility preservation.} In contrast to all GA baselines, MUTE achieves near-complete erasure ($\sim$1\% forget accuracy) while maintaining 55.0\% retain accuracy, representing a 2.75$\times$ improvement over the aggressive GA baseline at equivalent erasure levels. This demonstrates that targeting the language-agnostic layer is essential for effective multilingual unlearning without utility collapse.

\subsection{Sensitivity to Source Language Selection}
\label{appendix:source_lang_ablation}

To verify that MUTE's effectiveness is not tied to a specific choice of source languages, we repeat our main experiment with a different source language configuration. We use German, French, and Italian (DE, FR, IT) as source languages and treat the original source languages (EN, ES, PT) as held-out.

\begin{table}[h]
\centering
\caption{MUTE with alternative source languages (DE, FR, IT) on Llama-3.1-8B. $\dagger$ denotes source languages for this experiment. EN, ES, PT are now held-out languages.}
\label{tab:source_lang_ablation}
\resizebox{\textwidth}{!}{
\begin{tabular}{l|ccccccc|c|ccccccc|c}
\toprule
& \multicolumn{8}{c|}{\textbf{Forget Set Accuracy $\downarrow$}} & \multicolumn{8}{c}{\textbf{Retain Set Accuracy $\uparrow$}} \\
\textbf{Source Langs} & EN & ES & PT & DE & FR & HI & IT & Avg & EN & ES & PT & DE & FR & HI & IT & Avg \\
\midrule
Finetuned & 86.0\% & 74.0\% & 74.0\% & 68.0\% & 78.0\% & 64.0\% & 80.0\% & 75.0\% & 84.0\% & 66.0\% & 58.0\% & 58.0\% & 63.0\% & 23.0\% & 62.0\% & 59.0\% \\
\midrule
EN*, ES*, PT* & 2.0\% & 2.0\% & 1.0\% & 1.0\% & 1.0\% & 0.0\% & 2.0\% & 1.0\% & 77.0\% & 63.0\% & 56.0\% & 49.0\% & 57.0\% & 20.0\% & 61.0\% & 55.0\% \\
DE$\dagger$, FR$\dagger$, IT$\dagger$ & 1.0\% & 2.0\% & 1.0\% & 1.0\% & 1.0\% & 0.0\% & 2.0\% & 1.0\% & 75.0\% & 58.0\% & 54.0\% & 51.0\% & 58.0\% & 22.0\% & 58.0\% & 54.0\% \\
\bottomrule
\end{tabular}
}
\end{table}

Table~\ref{tab:source_lang_ablation} shows that MUTE achieves comparable performance regardless of source language selection. With DE/FR/IT as source languages, the method achieves $\sim$1\% forget accuracy across all languages (including the now held-out EN/ES/PT) while maintaining 54.0\% average retain accuracy. This is nearly identical to the original configuration (EN/ES/PT as source), confirming that MUTE's effectiveness stems from targeting the language-agnostic layer rather than from the specific choice of source languages.

\subsection{Mechanistic Verification on Llama-3.1}
\label{app:mechanistic}

Table~\ref{tab:mechanistic_results} presents the Logit Lens probing results, measuring the probability assigned to correct answers at intermediate layers.

\begin{table}[h]
\centering
\caption{\textbf{Mechanistic Verification (Logit Lens Recall) on Llama-3.1.} We probe internal representations to measure whether the model encodes the correct answer at intermediate layers. Shallow layers (L2) degrade Retain recall for held-out languages. Deep layers (L30) preserve Forget recall unchanged. Target layers (L9, L16) remove target knowledge while preserving the multilingual representation space. Asterisks (*) denote source languages.}
\label{tab:mechanistic_results}
\resizebox{\textwidth}{!}{
\begin{tabular}{l|cc|ccccc}
\toprule
\multirow{2}{*}{\textbf{Language}} & \textbf{Base} & \textbf{Finetuned} & \multicolumn{5}{c}{\textbf{Internal Recall by Intervention Layer}} \\
 & (Pre-trained) & (Target) & L2 (Shallow) & L4 (Shallow) & \textbf{L9 (Target)} & \textbf{L16 (Target)} & L30 (Deep) \\
\midrule
\multicolumn{8}{c}{\cellcolor{gray!10}\textbf{FORGET SET RECALL} (Lower is Better $\downarrow$)} \\
\midrule
English (en)* & 23.04\% & 29.39\% & 1.38\% & 1.40\% & \textbf{1.50\%} & \textbf{2.41\%} & 29.42\% \\
Spanish (es)* & 22.69\% & 28.18\% & 0.79\% & 0.81\% & \textbf{1.15\%} & \textbf{1.35\%} & 28.16\% \\
Portuguese (pt)* & 21.46\% & 27.68\% & 0.90\% & 0.92\% & \textbf{0.96\%} & \textbf{1.35\%} & 27.66\% \\
German (de) & 21.64\% & 27.44\% & 0.78\% & 0.79\% & \textbf{1.08\%} & \textbf{1.74\%} & 27.41\% \\
French (fr) & 21.96\% & 27.76\% & 0.77\% & 0.78\% & \textbf{0.81\%} & \textbf{1.29\%} & 27.77\% \\
Hindi (hi) & 21.88\% & 25.42\% & 0.53\% & 0.52\% & \textbf{1.18\%} & \textbf{1.11\%} & 25.40\% \\
Italian (it) & 21.52\% & 28.02\% & 0.78\% & 0.80\% & \textbf{0.87\%} & \textbf{1.70\%} & 28.01\% \\
\midrule
\multicolumn{8}{c}{\cellcolor{gray!10}\textbf{RETAIN SET RECALL} (Higher is Better $\uparrow$, closer to Finetuned)} \\
\midrule
English (en)* & 22.42\% & 32.30\% & 31.70\% & 32.12\% & \textbf{32.16\%} & \textbf{32.25\%} & 32.29\% \\
Spanish (es)* & 23.09\% & 31.28\% & 30.31\% & 31.05\% & \textbf{31.21\%} & \textbf{31.24\%} & 31.28\% \\
Portuguese (pt)* & 22.03\% & 31.39\% & 30.41\% & 31.11\% & \textbf{31.31\%} & \textbf{31.35\%} & 31.39\% \\
German (de) & 22.11\% & 30.71\% & 12.77\% & 28.08\% & \textbf{30.59\%} & \textbf{30.62\%} & 30.70\% \\
French (fr) & 22.73\% & 30.96\% & 23.54\% & 29.37\% & \textbf{30.70\%} & \textbf{30.85\%} & 30.95\% \\
Hindi (hi) & 25.06\% & 30.05\% & 3.47\% & 26.65\% & \textbf{29.64\%} & \textbf{29.77\%} & 30.04\% \\
Italian (it) & 22.89\% & 31.44\% & 12.63\% & 27.38\% & \textbf{31.32\%} & \textbf{31.30\%} & 31.44\% \\
\bottomrule
\end{tabular}
}
\end{table}

\subsection{Generalizability to Qwen-2.5-7B}
\label{app:qwen_results}

\begin{figure}[h]
    \centering
    \includegraphics[width=0.7\linewidth]{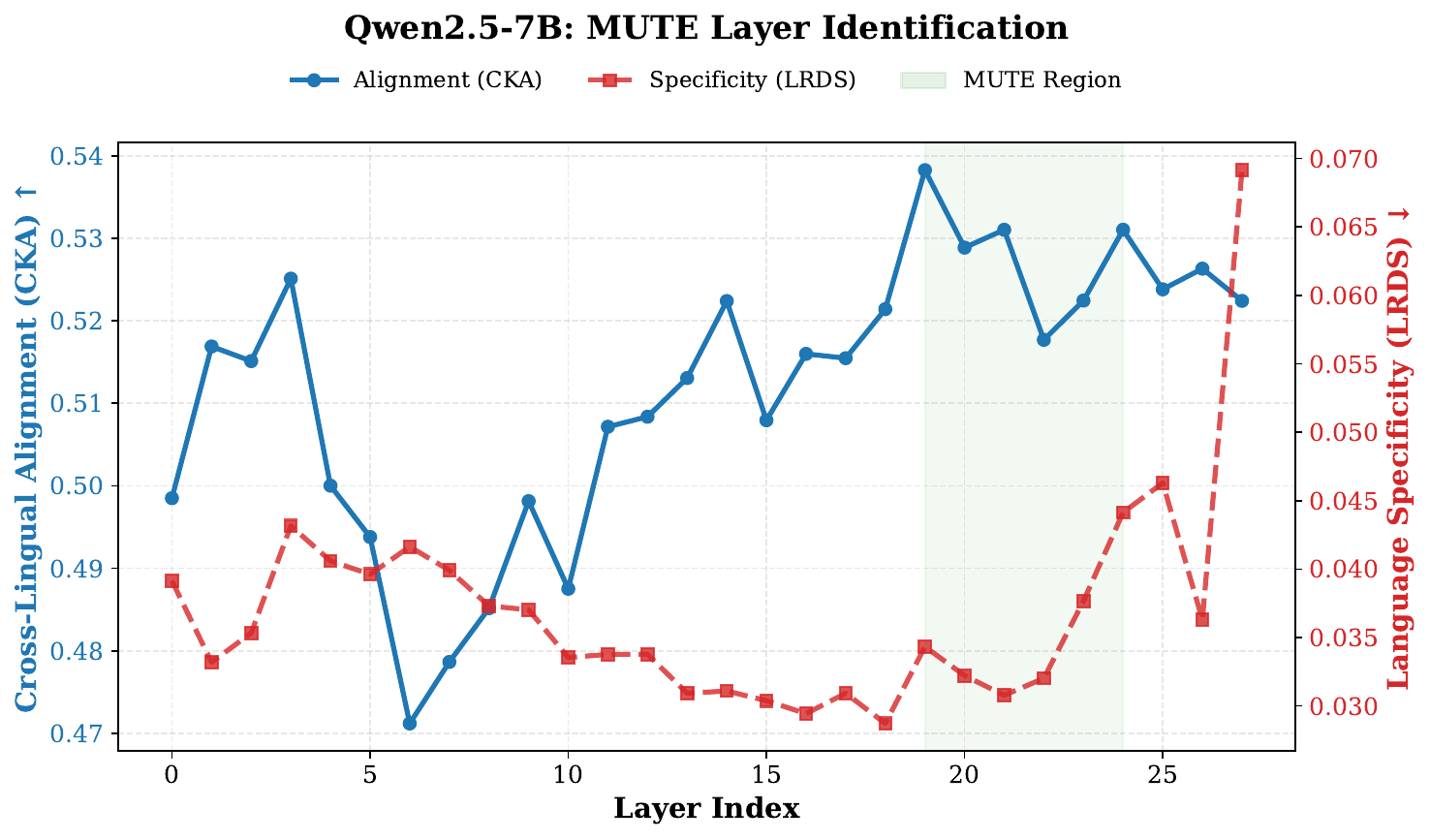}
    \caption{\textit{Identifying the language-agnostic region for Qwen-2.5-7B.} Blue line: multilingual CKA alignment (higher is better). Red line: LRDS (lower is more language-agnostic). Green shaded area: \textit{language-agnostic region} $\Lambda$, satisfying both $\text{CKA} > \tau_{\text{align}}$ and $\text{LRDS} < \tau_{\text{spec}}$, where thresholds are computed via Equation~\ref{eq:threshold}. We select Layer 19 as the optimal intervention point.}
    \label{fig:dual_metric_qwen}
\end{figure}

Figure~\ref{fig:dual_metric_qwen} presents the CKA-LRDS analysis for Qwen-2.5-7B, which identifies Layer 19 as the target layer. Table~\ref{tab:qwen_full} presents the corresponding unlearning results. At Layer 2 (shallow), we observe catastrophic forgetting in the Retain Set, with German accuracy dropping from 13.0\% to 3.0\%. Unlearning fails at Layer 23 (deep), as Forget accuracy remains high (English: 45.0\% vs. finetuned 52.0\%). At Layer 19 (target), the method achieves optimal erasure (26.0\%) while preserving utility across languages.

\begin{table}[h]
\centering
\caption{\textbf{Qwen-2.5-7B (RMU) across Layers.} Layer 19 represents the target layer. Shallow intervention (L2) causes Retain collapse in held-out languages, while deep intervention (L23) fails to erase knowledge. The target layer (L19) achieves optimal balance between erasure and utility preservation. Asterisks (*) denote source languages ($\mathcal{L}_{\text{src}}$).}
\label{tab:qwen_full}
\resizebox{\textwidth}{!}{
\begin{tabular}{l|cc|ccccc}
\toprule
\multirow{2}{*}{\textbf{Language}} & \textbf{Base} & \textbf{Finetuned} & \multicolumn{5}{c}{\textbf{Unlearned Accuracy by Intervention Layer}} \\
 & (Pre-trained) & (Target) & L2 (Shallow) & L12 & L15 & \textbf{L19 (Target)} & L23 (Deep) \\
\midrule
\multicolumn{8}{c}{\cellcolor{gray!10}\textbf{FORGET SET: High School Chemistry} (Lower is Better $\downarrow$)} \\
\midrule
English (en)* & 7.0\% & 52.0\% & 1.0\% & 3.0\% & 45.0\% & \textbf{26.0\%} & 45.0\% \\
Spanish (es)* & 7.0\% & 42.0\% & 0.0\% & 3.0\% & 29.0\% & \textbf{16.0\%} & 30.0\% \\
Portuguese (pt)* & 7.0\% & 46.0\% & 0.0\% & 1.0\% & 28.0\% & \textbf{15.0\%} & 32.0\% \\
Chinese (zh) & 7.0\% & 45.0\% & 1.0\% & 0.0\% & 30.0\% & \textbf{18.0\%} & 37.0\% \\
Arabic (ar) & 6.0\% & 35.0\% & 0.0\% & 0.0\% & 22.0\% & \textbf{15.0\%} & 25.0\% \\
German (de) & 5.0\% & 44.0\% & 0.0\% & 2.0\% & 27.0\% & \textbf{14.0\%} & 35.0\% \\
French (fr) & 4.0\% & 42.0\% & 0.0\% & 1.0\% & 31.0\% & \textbf{18.0\%} & 35.0\% \\
Hindi (hi) & 3.0\% & 31.0\% & 0.0\% & 2.0\% & 20.0\% & \textbf{11.0\%} & 28.0\% \\
Indonesian (id) & 10.0\% & 47.0\% & 0.0\% & 1.0\% & 31.0\% & \textbf{13.0\%} & 37.0\% \\
Italian (it) & 8.0\% & 49.0\% & 0.0\% & 2.0\% & 30.0\% & \textbf{18.0\%} & 37.0\% \\
Japanese (ja) & 6.0\% & 40.0\% & 0.0\% & 0.0\% & 24.0\% & \textbf{13.0\%} & 33.0\% \\
Korean (ko) & 6.0\% & 38.0\% & 1.0\% & 4.0\% & 21.0\% & \textbf{13.0\%} & 30.0\% \\
\midrule
\multicolumn{8}{c}{\cellcolor{gray!10}\textbf{RETAIN SET: History \& Law} (Higher is Better $\uparrow$, closer to Finetuned)} \\
\midrule
English (en)* & 7.0\% & 41.0\% & 12.0\% & 34.0\% & 37.0\% & \textbf{38.0\%} & 39.0\% \\
Spanish (es)* & 2.0\% & 21.0\% & 10.0\% & 17.0\% & 18.0\% & \textbf{16.0\%} & 20.0\% \\
Portuguese (pt)* & 2.0\% & 17.0\% & 10.0\% & 15.0\% & 18.0\% & \textbf{16.0\%} & 14.0\% \\
Chinese (zh) & 3.0\% & 12.0\% & 4.0\% & 12.0\% & 10.0\% & \textbf{12.0\%} & 10.0\% \\
Arabic (ar) & 2.0\% & 9.0\% & 4.0\% & 7.0\% & 8.0\% & \textbf{7.0\%} & 9.0\% \\
German (de) & 4.0\% & 13.0\% & 3.0\% & 8.0\% & 6.0\% & \textbf{10.0\%} & 12.0\% \\
French (fr) & 3.0\% & 16.0\% & 6.0\% & 15.0\% & 15.0\% & \textbf{15.0\%} & 19.0\% \\
Hindi (hi) & 0.0\% & 1.0\% & 0.0\% & 1.0\% & 1.0\% & \textbf{1.0\%} & 1.0\% \\
Indonesian (id) & 3.0\% & 16.0\% & 1.0\% & 8.0\% & 11.0\% & \textbf{13.0\%} & 13.0\% \\
Italian (it) & 3.0\% & 12.0\% & 4.0\% & 11.0\% & 15.0\% & \textbf{14.0\%} & 12.0\% \\
Japanese (ja) & 1.0\% & 10.0\% & 2.0\% & 5.0\% & 12.0\% & \textbf{9.0\%} & 10.0\% \\
Korean (ko) & 2.0\% & 10.0\% & 3.0\% & 7.0\% & 8.0\% & \textbf{8.0\%} & 9.0\% \\
\bottomrule
\end{tabular}
}
\end{table}

\subsection{Results on BLOOM-7b1}
\label{app:bloom_results}

\begin{figure}[h]
    \centering
    \includegraphics[width=0.7\linewidth]{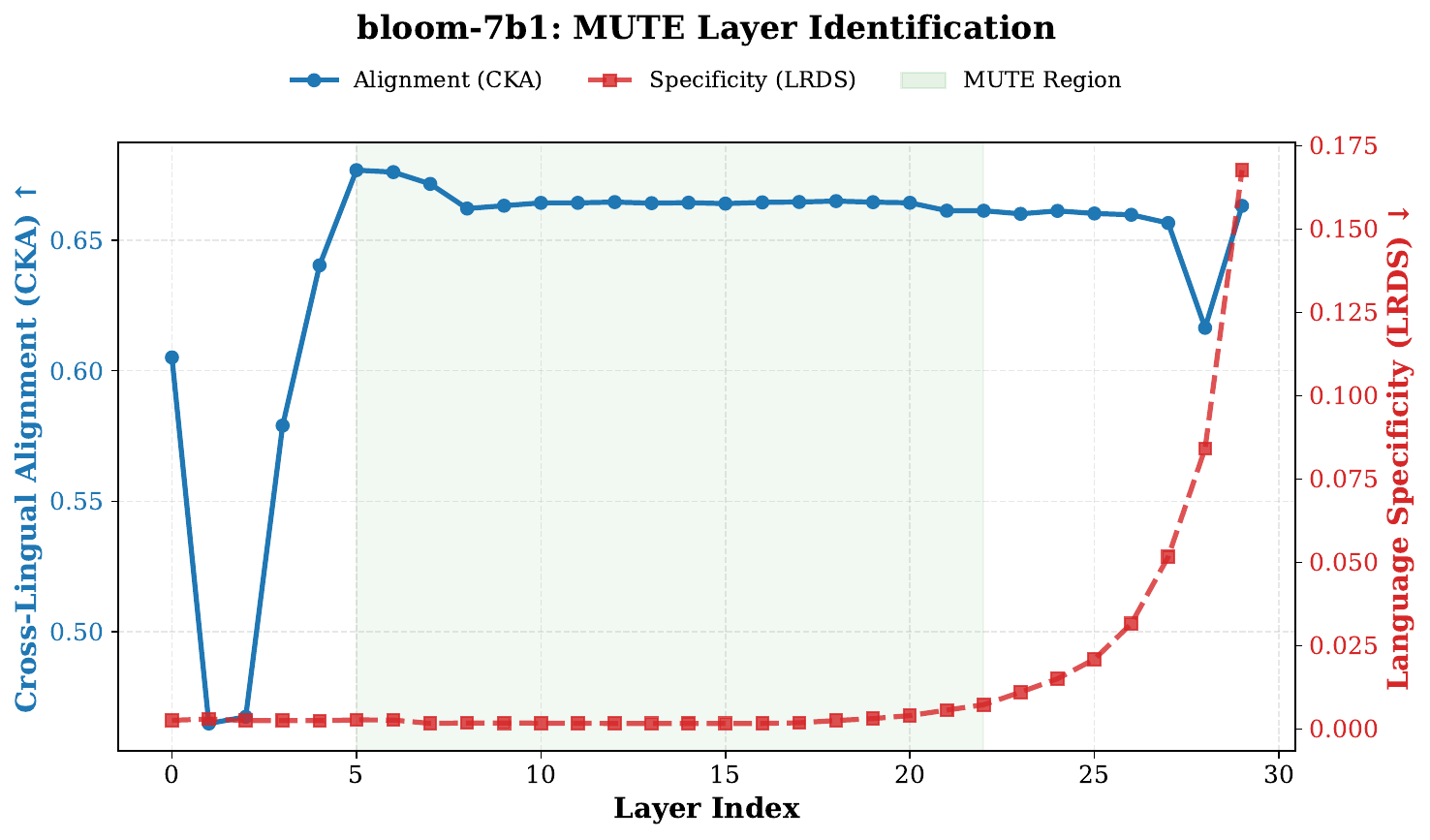}
    \caption{\textit{Identifying the language-agnostic region for BLOOM-7b1.} Blue line: multilingual CKA alignment (higher is better). Red line: LRDS (lower is more language-agnostic). Green shaded area: \textit{language-agnostic region} $\Lambda$, satisfying both $\text{CKA} > \tau_{\text{align}}$ and $\text{LRDS} < \tau_{\text{spec}}$, where thresholds are computed via Equation~\ref{eq:threshold}. We select Layer 5 as the optimal intervention point.}
    \label{fig:dual_metric_bloom}
\end{figure}

Figure~\ref{fig:dual_metric_bloom} presents the CKA-LRDS analysis for BLOOM-7b1, identifying Layer 5 as the target layer. Table~\ref{tab:bloom_full} presents the corresponding unlearning results. At Layer 29 (deep), unlearning fails entirely, with Forget accuracy remaining almost identical to the finetuned baseline (English: 18.0\% vs. 17.0\%). At Layer 5 (target), the method successfully erases knowledge (0.0\% Forget accuracy across all languages) while maintaining utility.

\begin{table}[h]
\centering
\caption{\textbf{BLOOM-7b1 (RMU) across Layers.} Layer 5 represents the target layer. Deep intervention (L29) fails to erase knowledge, with Forget accuracy unchanged. The target layer (L5) achieves complete erasure while preserving utility. Asterisks (*) denote source languages ($\mathcal{L}_{\text{src}}$).}
\label{tab:bloom_full}
\resizebox{\textwidth}{!}{
\begin{tabular}{l|cc|cccc}
\toprule
\multirow{2}{*}{\textbf{Language}} & \textbf{Base} & \textbf{Finetuned} & \multicolumn{4}{c}{\textbf{Unlearned Accuracy by Intervention Layer}} \\
 & (Pre-trained) & (Target) & L2 (Shallow) & \textbf{L5 (Target)} & L15 & L29 (Deep) \\
\midrule
\multicolumn{7}{c}{\cellcolor{gray!10}\textbf{FORGET SET: High School Chemistry} (Lower is Better $\downarrow$)} \\
\midrule
English (en)* & 1.0\% & 17.0\% & 1.0\% & \textbf{0.0\%} & 0.0\% & 18.0\% \\
Spanish (es)* & 2.0\% & 20.0\% & 0.0\% & \textbf{0.0\%} & 0.0\% & 16.0\% \\
Portuguese (pt)* & 3.0\% & 10.0\% & 0.0\% & \textbf{0.0\%} & 0.0\% & 10.0\% \\
Arabic (ar) & 1.0\% & 12.0\% & 0.0\% & \textbf{0.0\%} & 0.0\% & 8.0\% \\
French (fr) & 3.0\% & 15.0\% & 0.0\% & \textbf{0.0\%} & 0.0\% & 15.0\% \\
Hindi (hi) & 1.0\% & 14.0\% & 1.0\% & \textbf{0.0\%} & 0.0\% & 13.0\% \\
Indonesian (id) & 2.0\% & 13.0\% & 0.0\% & \textbf{0.0\%} & 0.0\% & 10.0\% \\
Chinese (zh) & 3.0\% & 13.0\% & 1.0\% & \textbf{0.0\%} & 0.0\% & 13.0\% \\
\midrule
\multicolumn{7}{c}{\cellcolor{gray!10}\textbf{RETAIN SET: History \& Law} (Higher is Better $\uparrow$, closer to Finetuned)} \\
\midrule
English (en)* & 1.0\% & 16.0\% & 15.0\% & \textbf{13.0\%} & 7.0\% & 16.0\% \\
Spanish (es)* & 2.0\% & 10.0\% & 6.0\% & \textbf{8.0\%} & 7.0\% & 10.0\% \\
Portuguese (pt)* & 1.0\% & 9.0\% & 8.0\% & \textbf{6.0\%} & 3.0\% & 9.0\% \\
Arabic (ar) & 1.0\% & 2.0\% & 1.0\% & \textbf{1.0\%} & 3.0\% & 2.0\% \\
French (fr) & 0.0\% & 9.0\% & 5.0\% & \textbf{9.0\%} & 4.0\% & 10.0\% \\
Hindi (hi) & 0.0\% & 1.0\% & 1.0\% & \textbf{0.0\%} & 0.0\% & 1.0\% \\
Indonesian (id) & 1.0\% & 7.0\% & 6.0\% & \textbf{7.0\%} & 2.0\% & 7.0\% \\
Chinese (zh) & 0.0\% & 8.0\% & 6.0\% & \textbf{7.0\%} & 6.0\% & 8.0\% \\
\bottomrule
\end{tabular}
}
\end{table}

\clearpage

\subsection{Results on SLUG (Llama-3.1)}
\label{app:slug_results}

Table~\ref{tab:slug_full} presents results using the SLUG algorithm. Unlike RMU and SimNPO, which target the early boundary of the language-agnostic region (Layer 9), SLUG requires richer semantic representations for accurate probing and thus targets deeper layers within the region (Section~\ref{sec:methodology}).

At Layer 3 (shallow), SLUG causes complete collapse, with both Forget and Retain accuracy dropping to 0\% across all languages, confirming that shallow interventions destroy fundamental multilingual representations. At Layers 18 and 20 (within the language-agnostic region), the method achieves partial erasure (Forget: 24.0\%--53.0\%) while preserving moderate utility (Retain: 7.0\%--58.0\%). Layer 20, identified as $l^*_{\text{SLUG}}$ by Equation~\ref{eq:slug_layer}, provides the best balance within this region. At Layer 30 (deep), unlearning fails entirely, with Forget accuracy remaining virtually identical to the finetuned baseline (English: 86.0\% $\to$ 88.0\%) while Retain performance is fully preserved.

SLUG's one-shot gradient update proves too aggressive when applied to early layers (L3, L9), causing utility loss. The results suggest that SLUG benefits from targeting the deeper boundary of the language-agnostic region, where representations are sufficiently abstract for effective concept removal while remaining language-agnostic. However, even at L20, the erasure-utility trade-off is less favorable than RMU or SimNPO, indicating that additional regularization may be needed for single-step unlearning methods.

\begin{table}[h]
\centering
\caption{\textbf{Llama-3.1 (SLUG) across Layers.} Following Equation~\ref{eq:slug_layer}, layers within the language-agnostic region (L18, L20) serve as target layers for activation-based methods. Shallow layers (L3, L9) cause complete collapse due to SLUG's aggressive single-step update. Deep layers (L30) fail to erase knowledge. Layers at the deeper boundary of the language-agnostic region (L18, L20) achieve partial erasure with moderate utility preservation. Asterisks (*) denote source languages ($\mathcal{L}_{\text{src}}$).}
\label{tab:slug_full}
\resizebox{\textwidth}{!}{
\begin{tabular}{l|cc|ccccc}
\toprule
\multirow{2}{*}{\textbf{Language}} & \textbf{Base} & \textbf{Finetuned} & \multicolumn{5}{c}{\textbf{Unlearned Accuracy by Intervention Layer}} \\
 & (Pre-trained) & (Target) & L3 (Shallow) & L9 & \textbf{L18 (Target)} & \textbf{L20 (Target)} & L30 (Deep) \\
\midrule
\multicolumn{8}{c}{\cellcolor{gray!10}\textbf{FORGET SET: High School Chemistry} (Lower is Better $\downarrow$)} \\
\midrule
English (en)* & 12.0\% & 86.0\% & 0.0\% & 0.0\% & \textbf{53.0\%} & \textbf{46.0\%} & 88.0\% \\
Spanish (es)* & 7.0\% & 74.0\% & 0.0\% & 0.0\% & \textbf{47.0\%} & \textbf{39.0\%} & 75.0\% \\
Portuguese (pt)* & 3.0\% & 74.0\% & 0.0\% & 0.0\% & \textbf{41.0\%} & \textbf{37.0\%} & 74.0\% \\
German (de) & 5.0\% & 68.0\% & 0.0\% & 0.0\% & \textbf{35.0\%} & \textbf{30.0\%} & 68.0\% \\
French (fr) & 5.0\% & 78.0\% & 0.0\% & 0.0\% & \textbf{40.0\%} & \textbf{37.0\%} & 76.0\% \\
Hindi (hi) & 6.0\% & 64.0\% & 0.0\% & 0.0\% & \textbf{30.0\%} & \textbf{24.0\%} & 62.0\% \\
Italian (it) & 5.0\% & 80.0\% & 0.0\% & 0.0\% & \textbf{47.0\%} & \textbf{35.0\%} & 82.0\% \\
\midrule
\multicolumn{8}{c}{\cellcolor{gray!10}\textbf{RETAIN SET: History \& Law} (Higher is Better $\uparrow$, closer to Finetuned)} \\
\midrule
English (en)* & 5.0\% & 84.0\% & 0.0\% & 0.0\% & \textbf{55.0\%} & \textbf{58.0\%} & 83.0\% \\
Spanish (es)* & 1.0\% & 66.0\% & 0.0\% & 0.0\% & \textbf{40.0\%} & \textbf{37.0\%} & 67.0\% \\
Portuguese (pt)* & 1.0\% & 58.0\% & 0.0\% & 0.0\% & \textbf{39.0\%} & \textbf{29.0\%} & 61.0\% \\
German (de) & 2.0\% & 58.0\% & 0.0\% & 0.0\% & \textbf{23.0\%} & \textbf{22.0\%} & 58.0\% \\
French (fr) & 1.0\% & 63.0\% & 0.0\% & 0.0\% & \textbf{37.0\%} & \textbf{37.0\%} & 63.0\% \\
Hindi (hi) & 0.0\% & 23.0\% & 0.0\% & 0.0\% & \textbf{12.0\%} & \textbf{7.0\%} & 24.0\% \\
Italian (it) & 2.0\% & 62.0\% & 0.0\% & 0.0\% & \textbf{44.0\%} & \textbf{30.0\%} & 62.0\% \\
\bottomrule
\end{tabular}
}
\end{table}

\subsection{Results on SimNPO (Llama-3.1)}
\label{app:simnpo_results}

Table~\ref{tab:simnpo_full} presents results using SimNPO. At Layer 30 (deep), unlearning fails with high Forget accuracy (81.0\%). At Layer 9 (target), the method achieves perfect erasure (0.0\%) with high retention (86.0\%), demonstrating the most favorable erasure-utility trade-off among the three algorithms evaluated.

\begin{table}[h]
\centering
\caption{\textbf{Llama-3.1 (SimNPO) across Layers.} Comparison of target and deep layers. Deep intervention (L30) fails to erase knowledge, with high Forget accuracy. The target layer (L9) achieves perfect erasure while maintaining strong utility across all languages. Asterisks (*) denote source languages ($\mathcal{L}_{\text{src}}$).}
\label{tab:simnpo_full}
\resizebox{\textwidth}{!}{
\begin{tabular}{l|cc|ccccc}
\toprule
\multirow{2}{*}{\textbf{Language}} & \textbf{Base} & \textbf{Finetuned} & \multicolumn{5}{c}{\textbf{Unlearned Accuracy by Intervention Layer}} \\
 & (Pre-trained) & (Target) & L2 (Shallow) & L3 & \textbf{L9 (Target)} & L18 & L30 (Deep) \\
\midrule
\multicolumn{8}{c}{\cellcolor{gray!10}\textbf{FORGET SET: High School Chemistry} (Lower is Better $\downarrow$)} \\
\midrule
English (en)* & 12.0\% & 86.0\% & 2.0\% & 1.0\% & \textbf{0.0\%} & 10.0\% & 81.0\% \\
Spanish (es)* & 7.0\% & 74.0\% & 1.0\% & 1.0\% & \textbf{0.0\%} & 5.0\% & 67.0\% \\
Portuguese (pt)* & 3.0\% & 74.0\% & 0.0\% & 1.0\% & \textbf{0.0\%} & 2.0\% & 67.0\% \\
German (de) & 5.0\% & 68.0\% & 0.0\% & 0.0\% & \textbf{0.0\%} & 2.0\% & 66.0\% \\
French (fr) & 5.0\% & 78.0\% & 0.0\% & 1.0\% & \textbf{0.0\%} & 3.0\% & 73.0\% \\
Hindi (hi) & 6.0\% & 64.0\% & 0.0\% & 0.0\% & \textbf{0.0\%} & 4.0\% & 57.0\% \\
Italian (it) & 5.0\% & 80.0\% & 2.0\% & 0.0\% & \textbf{0.0\%} & 2.0\% & 71.0\% \\
\midrule
\multicolumn{8}{c}{\cellcolor{gray!10}\textbf{RETAIN SET: History \& Law} (Higher is Better $\uparrow$, closer to Finetuned)} \\
\midrule
English (en)* & 5.0\% & 84.0\% & 79.0\% & 88.0\% & \textbf{86.0\%} & 83.0\% & 81.0\% \\
Spanish (es)* & 1.0\% & 66.0\% & 63.0\% & 71.0\% & \textbf{75.0\%} & 73.0\% & 66.0\% \\
Portuguese (pt)* & 1.0\% & 58.0\% & 59.0\% & 64.0\% & \textbf{63.0\%} & 62.0\% & 59.0\% \\
German (de) & 2.0\% & 58.0\% & 51.0\% & 55.0\% & \textbf{52.0\%} & 52.0\% & 57.0\% \\
French (fr) & 1.0\% & 63.0\% & 60.0\% & 64.0\% & \textbf{57.0\%} & 54.0\% & 62.0\% \\
Hindi (hi) & 0.0\% & 23.0\% & 18.0\% & 21.0\% & \textbf{19.0\%} & 23.0\% & 22.0\% \\
Italian (it) & 2.0\% & 62.0\% & 57.0\% & 60.0\% & \textbf{58.0\%} & 57.0\% & 60.0\% \\
\bottomrule
\end{tabular}
}
\end{table}


\clearpage
\section{Discussion}
\label{sec:discussion}

\noindent\textbf{Knowledge organization in multilingual models.} Our layer-wise analysis reveals a hierarchical organization of multilingual knowledge that has direct implications for unlearning. The catastrophic forgetting observed at shallow layers (e.g., Layer 2) indicates that these layers encode fundamental linguistic primitives shared across all languages. Intervening at this depth does not selectively remove knowledge but instead disrupts basic language processing capabilities.

In contrast, the intermediate layers identified by our CKA and LRDS metrics exhibit high multilingual alignment, where the same concept expressed in different languages converges to similar representations. This convergence explains the multilingual transfer observed in our experiments: modifying the shared semantic representation of a concept propagates the effect to all languages simultaneously.

The failure of deep-layer interventions (e.g., Layer 30) can be attributed to the fact that, at this depth, the target knowledge has already been retrieved and processed by earlier layers. Deep layers primarily handle language-specific token generation rather than core semantic representation; modifying them cannot erase knowledge that remains encoded upstream. Our Logit Lens analysis (Table~\ref{tab:mechanistic_results}) confirms this: optimization at these layers fails to modify the stored knowledge—internal activation probabilities remain virtually unchanged (e.g., English forget recall: 29.39\% $\rightarrow$ 29.42\%), explaining why behavioral metrics show no unlearning effect even in source languages.

\noindent\textbf{Distinguishing knowledge removal from output suppression.} A key finding of our work is the distinction between knowledge removal and output-level suppression. Standard unlearning methods that operate on final-layer logits risk creating what \citet{jia2025erasureillusionstresstestinggeneralization} term an Erasure Illusion: the model learns to suppress specific outputs rather than remove the underlying knowledge. Our deep-layer results demonstrate a different failure mode—not an illusion of erasure, but a complete failure to modify the target representations at all.

Our Logit Lens results demonstrate that interventions at the MUTE layer reduce internal activation probabilities of target concepts to near-zero across layers, indicating that the knowledge is removed from the model's computation path rather than merely hidden. This distinction has practical implications for safety-critical applications, where adversaries may attempt to recover supposedly unlearned knowledge through multilingual queries.

\end{document}